\documentclass[12pt]{iopart}

\pdfoutput = 1

\usepackage{iopams}
\usepackage{graphicx}
\usepackage[usenames]{color}
\usepackage[numbers]{natbib}
\usepackage{ulem}

\begin{document}

\def\RR{{\mathbb R}}
\def\cN{{\mathcal N}}
\def\PP{{\mathbb P}}
\newtheorem{remark}{Remark}

%*******************************
% Pade garde
%*******************************
\title[Detecting EEG Evoked Potential using a Wavelet Domain Linear Mixed Model]{Detecting Single-trial EEG Evoked Potential using a Wavelet Domain Linear Mixed Model: \\ \textit{Application to Error Potentials classification}}

\author{J Spinnato$^{1,2}$, M-C Roubaud$^1$, B Burle$^2$ and B Torr\'esani$^1$ }

\address{$^{1}$ Aix-Marseille Universit\'e, CNRS, Centrale Marseille, I2M, UMR 7373, 13453 Marseille, France}
\address{$^{2}$ Aix-Marseille Universit\'e, CNRS, LNC, UMR 7291, 13331 Marseille, France}
\ead{juliette.spinnato[AT]univ-amu.fr}

\begin{abstract}
\mbox{} \\
\textit{Objective.}
The main goal of this work is to develop a model for multi-sensor signals such as MEG or EEG signals, that accounts for the inter-trial variability, suitable for corresponding binary classification problems. An important constraint is that the model be simple enough to handle small size and unbalanced datasets, as often encountered in BCI type experiments.\\
%During the last decades, methods for EEG single-trial analysis have been largely developed, particularly for classification purpose. A key point is the study of the inter-trial variability that may carry important information regarding the underlying neural processes.\\
%In the present work we introduce a statistical model taking into account between trials %variability to extract the signal of interest from the background activity and to
%detect EEG evoked potentials.
\textit{Approach.} The method involves linear mixed effects statistical model, wavelet transform and spatial filtering, and aims at the characterization of localized discriminant features in multi-sensor signals. After discrete wavelet transform and spatial filtering, a projection onto the relevant wavelet and spatial channels subspaces is used for dimension reduction. The projected signals are then decomposed as the sum of a signal of interest (i.e. discriminant) and background noise, using a very simple Gaussian linear mixed model.\\
%The combination of these techniques provides a simple model, that indeed allows binary classification but also reconstruction of signals of interest.\\
% In particular, the construction of suitable pre-processing steps leads to both dimension reduction and
%simplication of the statistical model that accounts for the variability of signals of %interest and background noise.
\textit{Main results.} Thanks to the simplicity of the model, the corresponding parameter estimation problem is simplified. Robust estimates of class-covariance matrices are obtained from  small sample sizes and an effective Bayes plug-in classifier is derived.\\ The approach is applied to the detection of error potentials in multichannel EEG data, in a very unbalanced situation (detection of rare events). Classification results prove the relevance of the proposed approach in such a context.\\
\textit{Significance.} The combination of linear mixed model, wavelet transform and spatial filtering for EEG classification is, to the best of our knowledge, an original approach, which is proven to be effective. This paper improves on earlier results on similar problems, and the three main ingredients all play an important role.
\end{abstract}

%\pacs{?}
\submitto{Journal of Neural Engineering}

\maketitle

%*******************************
% Contenu
%*******************************
% INTRODUCTION - JNE

\section{Introduction}
\label{se:intro}

Electro- and Magneto-encephalography (respectively, EEG and MEG) are of the rare techniques allowing non-invasive brain investigation with an excellent temporal resolution and, under some conditions, a fairly good spatial one. One main limitation, however, is that extracting the brain activity of interest from background activity and noise usually requires averaging a large number of repetitions of the signal recorded in the ``same'' condition (for example, averaging several epochs of signal following the same repeated stimulus). Such an averaging, however, distorts the signal and prevents precise investigation of the dynamics of the underlying processes. Indeed, it has long been known that averaging has non-linear impact on the latencies estimation \cite{callaway:84,knuth2006,burle:08b}. For example, the latency of the onset of an activity measured on the average largely underestimates the real mean of the individual onsets \cite{smulders:96,meyer:88,kukleta:01}, making its use problematic for direct comparison with chronometric variables such as Reaction Time (RT) \cite{burle:08b,meyer:88}. The averaging process also prevents from analyzing learning and/or adaptation effects across trials repetitions \cite{quiroga:02}. More generally averaging trials eliminates the signal of interest variability, although the latter contains important information that can be useful in various contexts, from signal interpretation to classification.

For these reasons, methods for single-trial EEG analysis and classification have been developed during the last decades. %, among others
To reduce the impact of noise and background activity in single-trial analysis, common strategies  have been to extract more elementary parts concentrating the signal of interest, either in time, in frequency, or in space. Such approaches include, among others, selection of  time domains, frequency filtering, wavelet decomposition \cite{quiroga:00,wang:07}, adaptive basis selection \cite{Vautrin2009,Barbieri13optimal}, matching pursuit \cite{Durka07matching,Benar09consensus} or blind source separation \cite{jung:01}.
% and many others
 In those approaches, one aims at simply getting rid of the variability induced by the background activity and the noise, hoping that the remaining variability will only be attributable to the signal of interest.

% In this paper we are investigating modelling approaches mainly based on class-covariance definition and estimation.
Alternative strategies can be  based on explicit signal modelling. For example the Linear Discriminant Analysis (LDA) which is  one of the most popular classifiers for EEG single-trials detection (and for brain computer interface - BCI - applications, see \cite{Lotte2007}) can be interpreted in terms of very simple Gaussian mixture model. In this setting LDA assumes a class-independant covariance matrix. In a recent overview~\cite{blankertz:11}, Blankertz and coworkers proposed a regularized Gaussian mixture model for event related potentials (ERPs). In the latter,  the single-trial is written as the sum of a signal of interest which is approximated as  constant over trials and a random background activity modelled as a Gaussian noise. The inter-trial variability  is not taken into account.
 The class-covariance matrices are assumed to be equal, which leads to a very simple detection algorithm that turns out to be very efficient for binary classification tasks. 
 
However when this equality assumption is not valid,   the problem becomes more complex, especially when the two classes are unbalanced.
% In the present work, we address the binary classification problem where one of the two classes is much larger than the other one. 
In that context, standard classifiers, such as LDA, may fail as the estimated common covariance matrix is largely determined by the majority  class~\cite{Hand2003,Xie2006}. Taking into account the difference of the class-covariance matrices leads to a quadratic classification rule (QDA) and requires the estimation of a covariance matrix for each class.  This leads to a robustness  problem in the estimates, the more so when the size of the minority class is small.

% And it is well known that  when the size of the datasets are not enough to estimate precisely the two class-covariance matrices, the results of the LDA classifier are better than those of the QDA one even in the case where the variability is different in the two classes.

In BCI and more generally in EEG experiments, such  an unbalanced situation is not unfrequent. For example in BCI, the P300 speller protocol naturally generates two unbalanced classes \cite{Farwell1988}. Classical P300 spellers, with a $6\times 6$ matrix containing all letters and characters,  yield an unbalanced datasets composed of $1/6$ of ERP signals and of $5/6$ of noERP. 
Another example of unbalanced classes can be found in RT tasks in which the participant has to respond as fast as possible to the appearance of predetermined stimulation. For example, standard experimental psychology protocols make use of biased probabilities across experimental conditions (see \emph{e.g.} \cite{Posner1980}). Besides manipulated factors, biased probability might also be the result of participants behavior, such as errors which are typically much lower than correct trials.

% However, two situations may be considered: 
%In such unbalanced situations, two cases must be considered:
%if the two class %co
%variances are identical, then unbalancedness will not affect the classifier results. If the
%assumption of equal %co
%variances does not hold, both modelling and linear classifiers tend to be biased toward the majority class
%and classification results appear to be degraded~\cite{Hand2003,Xie2006}.{\color{blue}Est ce le bon endroit pour parler de covariances ? parler plutôt de variabilit\'e }\textcolor{red}{D'accord, rester sur les variances est plus clair}

In this paper, we target more specifically situations where the class-covariance matrices differ and available datasets are unbalanced and of limited size.  We propose  a wavelet domain Gaussian linear mixed model (termed LMM in the following)
for the binary signal classification problem. 
The model expresses each single-trial  as a sum of a class-dependent signal of interest and a  background activity as in~\cite{Huang2008,Huang2009}. The signal of interest is modelled as a multivariate Gaussian vector whose covariance matrix, that  describes the inter-trial variability,  depends on a user-specified design matrix. Both mean and design matrix are class-dependent. The background signal is modelled as a class-independent Gaussian white noise. The resulting model is characterized by a remarkably small number of parameters. 

The application context of the current paper is the detection of evoked potentials in M/EEG signals. 
The above described model turns out to be relevant for such signals when suitable preprocessings are performed, namely wavelet based time decorrelation, spatial filtering and corresponding dimension reductions. The procedure is more specifically applied to the problem of error negativity (ErrP) detection and analysis. Corresponding classification results compare very favorably with standard linear classifiers for small sample size datasets. 

\medskip

%Both modelling and discriminant methods are commonly used in M/EEG applications. In the present work we combine those two approaches
%in order to build a solid procedure for EEG single-trial analysis and classification, based on a LMM.
One of the main contributions of this work is a model for subject specific M/EEG signals, that belongs to the family of linear mixed models. Mixed models offer a rich and flexible framework for describing and quantifying variability and produce robust estimates of class-covariance matrices from small datasets.  
Such models have been considered in EEG applications in many different contexts. Let us for example mention~\cite{Baayen2008} where an introduction to mixed model is provided and possible applications in psychology and neuroimaging are discussed.
Mixed models have also recently  been used by other authors in bayesian framework for electrophysiology applications. In~\cite{Davidson2009}, a functional mixed  model is introduced for the purpose of regression analysis of ERP's data and where the model is combined with discrete wavelet transform and sparsity-inducing prior distribution.
In~\cite{Fazli2011} a subject-independent classifier is proposed  using a $\ell_1$-penalized linear mixed model. Within-subject and  between-subject variabilities are modelled through random and fixed terms and the method is applied to BCI achieving  subject-to-subject transfer.
The present work, inspired on the work of Huang and coworkers~\cite{Huang2008,Huang2009}, focuses on a subject specific model taking into account the variability across repetitions of the same experiment. The model proposed here is constructed directly in wavelet domain instead of time domain, and also differs from \cite{Huang2008,Huang2009} in both fixed and random parts. In our case the fixed part corresponds to the class-average whereas they used projection onto the first principal components. The modelling of inter-trial variability is different too. All  together this results in a simplified  Gaussian linear mixed model with less parameters  to estimate. Moreover the application of the model to the binary classification problem is different. In our work we directly exploit the estimates of class-covariance matrices in a Bayes classifier  while  Huang~\etal{}~\cite{Huang2008,Huang2009} use a likelihood test based on the predicted single-trial signals. Let us stress that we only address and model here the inter-trial variability. Even though the dataset used for validation involves several subjects, they are all treated independently and we do not attempt to model the inter-subject variability.
%Our approach aims at explicitely modelling single-trial variability and provide a simplified Gaussian LMM for the binary classification problem.}
\smallskip

This paper is organized as follows. A short overview for linear and quadratic discriminant methods for unbalanced data is given in~\sref{se:classif_unbal}. The model and the statistical procedure are described in~\sref{se:methods}
while~\sref{se:application} is devoted to the application to ErrP data. Discussion and conclusions are given
in~\sref{se:discussion} and~\sref{se:conclusion} respectively.
%
%

%\pagebreak

\section{A short overview of linear and quadratic discriminant analysis for unbalanced data}
\label{se:classif_unbal}

%**********************************
%In this section the unbalanced situation is addressed for discriminant analysis-based classifiers. 
%We particularly provide a brief review on linear and quadratic discriminant analysis and we introduce 
%some estimation issues associated with the case of non-equal and relatively small datasets.
%Then, a solution is investigated by considering the introduction of diagonally dominant covariance matrices.
We briefly discuss here the main issues encountered when using linear and quadratic discriminant analysis for unbalanced and small size datasets, and describe some classical solutions that will be of interest for this work.

%% LDA and QDA
\subsection{Linear and Quadratic discriminant analysis}
\label{sse:LDAQDA}
Let us take the probabilistic point of view, and derive Linear Discriminant Analysis (LDA) and Quadratic Discriminant Analysis (QDA) from the decision problem in the case of a Gaussian mixture, see \emph{e.g.}~\cite{Hastie2009}.
Observations are considered as  multivariate Gaussian draws with respective prior probabilities $p^c$ (with $\sum_c p^c=1$) and probability density function (pdf):
\begin{equation}
f^c(x) = \frac{1}{(2\pi)^{d/2}\vert\Sigma^c\vert^{1/2}} \exp\left[ -\frac1{2}(x-\mu^c)'(\Sigma^c)^{-1}(x-\mu^c)\right]\ ,
\end{equation} 
where $\mu^c \in \mathbb{R}^d$ is the class-mean, $\Sigma^c \in \mathbb{R}^{d\times d}$ is the class-covariance matrix (assumed to be invertible) and $\vert\Sigma^c\vert$ its determinant.

This leads to the discriminant function, defined for each class $c$:
\begin{equation}
\label{eq:discrimfct}
\delta^c(x) = -\frac{1}{2} \log \vert\Sigma^c\vert -\frac{1}{2} (x-\mu^c)'(\Sigma^c)^{-1}(x-\mu^c) + \log p^c\ .
\end{equation}
Let $x_i$ denote the $i$-th sample to classify. $x_i$ a $d$-dimensional vector. 
The decision rule is given by assigning $x_i$ to the class 
with the maximal $\delta^c(x_i)$. In the general case,
the class-covariances $\Sigma^c$ are different, and the decision should be based on quadratic functions of $x_i$ (thus the name QDA). When the class-covariances are assumed to be equal ($\Sigma^c = \Sigma$ for all $c$) the decision is based on linear functions of $x_i$ (LDA).

In practice the parameters of the Gaussian distribution in each class  are unknown and must be estimated, which is problematic in the case of small size and/or high dimensional datasets. We address this problem below, focusing on binary classification ($c\in \{0,1\}$) and unbalanced datasets. We will denote by $N^0$ and $N^1$ the corresponding sample sizes of datasets from which the estimation is done, and set $N=N^0+N^1$.
$\hat\Sigma^c$ and  $\hat\Sigma$ will denote respectively the sample estimates of class-covariance matrix $\Sigma^c$ and common covariance matrix $\Sigma$  defined as follows: 

\begin{equation}
\label{sampleclass-cov}
\hat\Sigma^c= \frac{1}{N^c-1}\sum_{i=1}^{N^c} (x_i-\bar{x}^c)(x_i-\bar{x}^c)' \ ,\quad\mbox{where} \quad  \bar{x}^c= \frac{1}{N^c}\sum_{i=1}^{N^c} x_i \,,
\end{equation}

\begin{equation}
\label{samplecov}
\hat\Sigma= \frac{N^0}{N} \hat\Sigma^0 + \frac{N^1}{N} \hat\Sigma^1\,.
\end{equation}

\smallskip

Two cases are to be considered.
\paragraph{Case 1: Equal class-covariance matrices.}
When $\Sigma^0=\Sigma^1=\Sigma$, the decision is based upon the sign of $w'x_i$, where $w = \Sigma^{-1}(\mu^0 - \mu^1)$.
The quality of the decision relies heavily on the quality of the estimation of the covariance matrix $\Sigma$ and its inverse, which may turn out to be poor for small datasets and/or in high-dimensional situations. In addition, low quality estimate for $\Sigma$ may lead to invertibility problems, which makes the classifier ill-defined. Solutions to such problems are discussed below.

Let us note that for equal class-covariance matrices, unbalancedness of the datasets does not introduce additional difficulty as $\Sigma$ is estimated from the complete dataset.

\paragraph{Case 2: Unequal class-covariance matrices.}
In this case, two covariance matrices must be estimated, which amplifies the above mentioned difficulty, particularly when the dataset is small and unbalanced enough so that at least one of the two covariance matrices is poorly estimated. This is why LDA is generally preferred, even when the true class-covariances are different.
However, when the datasets are unbalanced, the estimation tends to focus on the prevalent class and to ignore the rare one: the covariance estimate $\hat{\Sigma}$ given in (\ref{samplecov}) is dominated by the majority class, so that ${\hat\Sigma} \approx {\hat\Sigma}^0$ (in the case where $N^1 \ll N^0$).

In such a situation, a common solution consists in re-balancing datasets using sampling methods~\cite{Xie2006,Xue2008,Weiss2004}. 
Over-sampling increases the size of the minority class (without information gain) while under-sampling reduces the size of the majority class (which may result in prejudicial information loss).

%% Diagonally dominant matrices
\subsection{The case of diagonally dominant covariance matrices}
We stick here to binary classification problem, in the small sample size ($d\gg N$) and unbalanced ($N^1\ll N^0$) situation. In such case, estimated covariance matrices are often non-invertible, and some additional assumptions or regularization are needed. We review a few solutions that will be of interest for us, that ensure invertibility by enforcing diagonal dominance of the covariance matrices.

\begin{enumerate}
\item
{\em Diagonal covariance }.
If data are assumed to be decorrelated, the covariance matrix is diagonal and only the $d$ variances are to be estimated (leading to Diagonal LDA, DLDA for short). 
 Under the multivariate Gaussian law, this assumption corresponds to independence assumption.  However it is known that even for correlated data (which is the general situation) better classification results are often obtained when correlations are ignored, thus replacing $\hat \Sigma$ by its diagonal (leading to the so-called naive Bayes classifier). This is supported by asymptotic arguments ($d \gg n$) as well as experimental results (see \emph{e.g.} \cite{Bickel2004} and references therein). In particular if the off-diagonal elements of the covariance matrix are expected to be  nearly zero, it is usual to estimate them by zero.  Otherwise in some situations, a decorrelating transformation can be found that enforces the diagonal dominance of the covariance matrix (see section~\ref{sse:preprocessing}). 

%The diagonal approach given
%in~\ref{eq:diagLDA} can be extended to the quadratic situation~\cite{Dudoit2002} 
%(diagonal QDA):  

The diagonal assumption as well as the above considerations can be extended to the quadratic situation (leading to the diagonal QDA, DQDA for short).

\item {\em Diagonal dominance enforcing regularization}. 
Singularity issues can be overcome by shrinking the sample covariance matrix towards a multiple of the identity matrix~\cite{Friedman1988}. A particular trace preserving shrinkage approach has been proposed in~\cite{blankertz:11} in the context of EEG single-trial analysis. 
The corresponding estimate is a weighted average of  the sample covariance matrix and  the identity matrix 
\begin{equation}
\hat{\Sigma}(\gamma) = (1-\gamma)\hat{\Sigma} + \gamma\frac{\tr(\hat{\Sigma})}{d}  \ \mathbf{I}_{d}\ ,
\end{equation}
where $\gamma \in \left[0,1\right]$ is a shrinkage parameter and $\tr(\hat{\Sigma})$ is the trace of the sample covariance matrix, i.e. the sum of its diagonal elements. The standard practice is to find the optimal $\gamma$ by cross-validation. Blankertz~\etal give a simple heuristic estimate with satisfactory results on EEG data~\cite{blankertz:11}. This method will be called Regularized LDA (RDA for short).

\item  {\em Explicit covariance modelling}. 
When  prior information on data is avalaible, explicit models for covariance matrices can sometimes be proposed to reduce the number of parameters to estimate. We describe below an example where  a suitable transformation combined with a linear mixed model leads to diagonal dominant class-covariance matrices characterized by $\left( 2 + 2\times d\right) $ parameters only.
    
\end{enumerate}

%\pagebreak
\section{Method}
\label{se:methods}
We propose a classification method for multisensor signals to discriminate between two experimental conditions in view of application to EEG and BCI. The signals are  recorded over $M$ sensors in a fixed time-period with $T$ samples. For a given participant, we obtain $N^c$ trials per class, $c\in \{0,1\}$ being the class label, with each trial taking the form of $M$ time series of $T$ samples each.

%\footnote{those $T$ values can be the mere time course of the recorded signal,  or any %linear transformations of it (see below)}.

Our procedure is based on three steps. The first step is  preprocessing: transformations are applied to obtain diagonally dominant sample class-covariance matrices and the temporal and spatial dimensions are reduced. The second step  models variability using a Gaussian linear mixed model to estimate the class-covariance matrices. The third step is  classification.

%%%
%Preprocessing procedure to reduce spatial
%  and temporal dimensions in a discriminant fashion. LDA-based method on sensors defines
%  the $J$ most discriminative channels. Using DWT the signal of interest is
%  concentrated in a small number of wavelet coefficients. As the first levels of
%  decomposition correspond to high frequencies, low-pass filtering significantly reduces the
%  temporal information  in the wavelet domain.
%%%%%

%Parler des dimensions du signal (temps/electrodes).
%
%On cherche  se ramener  une matrice de cov  dominante diag en utilisant un
%pr-traitement et ensuite ou l'estime par le mod mixte.

%***************************
\subsection{Decorrelation and dimension reduction}
\label{sse:preprocessing}

Raw recorded signals are known to be strongly correlated both across sensors and time points. Data are thus preprocessed  in order to reduce dimensions and enforce the diagonal dominance of class-covariance matrices.

%\begin{figure}
%\centering
%%\includegraphics[width=0.8\textwidth]{figures/Graph_Pretreatment.pdf}
%\includegraphics[width=0.6\textwidth]{figures/Spatial_Temporal_Reduction}
%\caption{\label{Preprocessing}  Preprocessing for decorrelation and spatial and temporal dimensions reduction.}

%% Time-domain decorrelation
\subsubsection{Time-domain decorrelation using DWT.}
\label{ssse:wavelets}
Signals are recorded as time-series, with a given sampling frequency. However,
samples are expected to be highly correlated, should they correspond to the
background activity or to the signal of interest. Therefore, it is important to
reduce the correlations, which we do through a linear transform, by replacing
the time samples with the coefficients of the signals expansion with respect to
a suitably chosen basis.  
In this work, we limit ourselves to a wavelet basis, generated by
shifting and rescaling a generic waveform $\psi$. More precisely, in
the framework of the so-called {\it multiresolution analysis}, it is
possible to find pairs of functions $(\phi,\psi)$ such that suitably rescaled
and shifted copies of these form orthonormal bases of signal spaces of interest.
Without going into mathematical details, for which we refer to \emph{e.g.}~\cite{Daubechies92ten,Mallat08wavelet,Vetterli95wavelets} for detailed accounts,
and also to~\cite{Hubbard98world} for a pedestrian introduction, this yields
multiscale signal decompositions as follows. Given a reference (integer) scale index $m_0$, any signal in the signal space can be uniquely expanded as time-shifted copies of the scaling function at scale $2^{m_0}$ and time-shifted copies of the wavelet, rescaled at smaller scales $2^m, m\le m_0$ (also called {\it levels}). This is expressed mathematically in the form
\begin{equation*}
f(t) = \sum_n s_{m_0,n} 2^{-m_0/2}\phi(2^{-m}t-n) +
\sum_{m\le m_0}\sum_n d_{m,n} 2^{-m_0/2}\psi(2^{-m}t-n)\ .
\end{equation*}
The coefficients that enter such an expansion are called respectively
{\it scaling coefficients} (for the coefficients $s_{m_0,n}$) and {\it wavelet
coefficients } (coefficients $d_{m,n}$) and are easily computed as inner products
of the signal $f$ with the corresponding basis functions. The numerical computation
of these coefficients is performed using dedicated fast algorithms
(see~\cite{Vetterli95wavelets}), which leads to the so-called DWT (for discrete
wavelet transform). In what follows, we will gather scaling and wavelet
coefficients of a signal in a unique vector of {\it multiscale coefficients}.

\medskip
Wavelets are introduced here because of their ability to decorrelate signals. It has been found empirically in many application domains (such as image processing) that the nonzero (or significant) multiscale coefficients of a correlated signal are often far less correlated than the signal, and that the number of significant coefficients is therefore much smaller than the signal length. This turns out to be the case for the EEG signal considered here. Therefore, a (time-domain) dimension reduction can be performed by moving to the multiscale domain using DWT, and setting to zero irrelevant multiscale coefficients, namely those coefficients that are  numerically negligible for all channels. The resulting coefficients are therefore weakly correlated (and the corresponding covariance matrix is strongly diagonal dominant). It is worth noticing that wavelets only provide approximate decorrelation. Alternatives to wavelets have been proposed for improving the decorrelation by seeking optimally decorrelating basis (see~\cite{Wickerhauser96adapted} for detailed account of the best basis approach and ~\cite{Vautrin2009,Barbieri13optimal} for the adaptation to EEG signals).
These techniques are adaptive and, as a consequence, the decomposition basis is signal dependent, which is not suitable for the model we are using here. 
%We will therefore not consider such techniques here and stick to standard wavelet decompositions.

In addition to dimension reduction, a level-dependent rescaling is performed on the retained multiscale coefficients to correct for variability differences across scales. More precisely, at each scale, coefficients are normalized so that they all have the same variance.

As a result of this time (or time-scale) domain processing, each trial $i$ gives rise to a multiscale coefficient matrix $X_i^c \in \RR^{M\times K}$ ($c$ being the class label), where $K$ is the number of retained multiscale coefficients and $M$ being the number of sensors.

\subsubsection{Spatial filtering.}
\label{sss:SpatialFilt}
The considered signals are multisensor signals, with low spatial resolution and large spatial correlations. These can be reduced using spatial filtering. In single-trial analysis, spatial filters are used to improve
signal-to-noise ratio while reducing both EEG data complexity and spatial dimension.
Various spatial filter constructions have been proposed, depending on EEG application (see~\cite{Parra2005,McFarland1997} for reviews). 

In the present study, we rely on a matrix-based technique inspired by Fisher's
linear discriminant method~\cite{Fisher1936}, as done in~\cite{Huang2009}.  The main purpose is to identify projections in the sensor space that keep the classes as separated as possible while minimizing the variance within classes.
We introduce the between-class and the within-class  covariance matrices $S_B$ and $S_W$, respectively
\begin{eqnarray}
& S_B &= \frac{1}{NT} \sum_{c=0}^1 N^c \left(\overline{X}^c - \overline{X} \right)\left(\overline{X}^c - \overline{X} \right)^\prime , \\
& S_W &= \frac{1}{NT} \sum_{c=0}^1 \sum_{i=1}^{N^c} \left(X_i^c - \overline{X}^c \right)\left(X_i^c - \overline{X}^c \right)^\prime \ ,
\end{eqnarray}
where $\overline{X}^c \in \mathbb{R}^{M\times K}$ and $\overline{X} \in \mathbb{R}^{M\times K}$ are the trial-average in class $c$ and the total trial-average matrices. Here, $N=N^0+N^1$ is the total number of trials. Notice that $S_B$ and $S_W$ are both $M\times M$ non-negative definite matrices.

The method seeks to identify the most discriminant linear combinations of sensors by optimizing the following 
Fisher criterion, sequentially under orthogonality constraints between vectors:
\begin{equation}
\max_{u \in \mathbb{R}^{M}}  \frac{u' S_B u}{u' S_W u}.
\end{equation}

The solution is obtained by computing the eigenvectors decomposition of the matrix $S_W^{-1} S_B$  (assuming that $S_W$ is invertible). 
The eigenvectors are sorted  by decreasing discriminating power (eigenvalue), out of which the first $J$ are selected. The number $J$ is chosen according to a criterion based on the cumulative 
percentage of eigenvalues. 
In most situations, $J \ll M$, which yields a significant spatial dimension reduction. The $J$  uncorrelated linear combinations of the $M$ sensors will be called {\it channels} in the following.

Let $U=\left[u_1,\ldots,u_J\right]$ denote the matrix whose columns are the selected eigenvectors. Each single-trial signal $X_i^c$ is projected onto the selected subspace spanned by these $J$ eigenvectors as follows
\begin{equation}
\label{YChan}
Y_i^c = U^\prime X_i^c\, ,
\end{equation}
where each row $j$ of the matrix $Y_i^c \in \mathbb{R}^{J\times K}$ consists in the wavelet and scaling function coefficients of the $j$-th channel for single-trial~$i$.
Each column of $U$, denoted $u_j$, can be referred to as \textit{spatial filters} (see~\cite{McFarland1997,blankertz08,blankertz:11,Parra2005}).

%Notice that in this process, the background activity, which does not differ
%significantly in the two classes is spread out into the $M-J$ non-selected channels, leaving only a {\it residual
%background activity}.
%
%Lien avec le quotient de Rayleigh (pas mal de biblio basee sur cette idee pour le filtrage spatial en BCI).

%%***************************
%\subsubsection{Testing homogeneity of class-covariance matrices }
%
%\begin{figure}[h]
%	\centering
%		\includegraphics[width=1\textwidth]{../../ArticleNeuralEngin/Figures_JNE/CovMatrix_KJ.pdf}
%	\caption{\label{fig:Cov_KJ} Representation of the class covariance matrices after decorrelation and features extraction. 
%	The Error-class covariance $\Sigma^{Err}$ and the Correct-class covariance $\Sigma^{PC}$ are square matrices of dimension $KJ\times KJ$ where 
%	$K=24$ is the number of wavelet coefficients and $J=2$ is the number of channels. The two covariances are significantly different (Box's M Test for homogeneity of covariances).  }
%\end{figure}
%
%\begin{table}[h]
%\caption{\label{tab:MBoxTest} Box's M test for equality of the two covariance matrices $\Sigma^{Err}$ and $\Sigma^{PC}$.}
%\begin{indented}
%\item[] 
%\begin{tabular}{@{} l l l l l l}
%\br 
%MBox & Chi-sqr & g & p & df & p-value \\
%\mr
%3273.0507 & 2859.7354 & 2 & 48 & 1176 & 0.0000 \\
%\br
%\end{tabular}
%\end{indented}
%\end{table}

%***************************
\subsection{Model setup}
\label{se:modelling}

In this section a single-trial analysis is proposed, in which the variability is modelled through a Gaussian linear mixed model~\cite{Rao1988,McCulloch2008}.

\subsubsection{Background and definition.}
\label{sss:model.definition}

 Given multisensor and multi-trial signals $f_{i,s}^c(t)$,  with $i$ the trial index and $s$
the sensor index, that are labelled by a class index $c$, we assume that such
signals can be modelled as a sum of two independent components:

\begin{itemize}
\item
A background activity, assumed to be stationary (in the sense of weakly stationary random processes, namely the two first order moments are translation invariant)
and correlated (see \emph{e.g.}~\cite{koopmans1995}),
\item
Event related components (also called signals of interest), thus
intrinsically non-stationary, characterized by a {\it class index} which labels the
cerebral response to the events.
The signal of interest is further modelled as the sum of a fixed component,
common to all the trials in the class, and a random trial dependent component
(the so-called {\it trial random effect}). 
\end{itemize}
In what follows, the signals that will be modelled in such a way are the multi-channel multiscale (wavelet and scaling) coefficient arrays $Y_i^c\in \RR^{J\times K}$  resulting from the preprocessing. We will denote by $y_i^c \in \RR^{JK}$ the vector obtained by concatenation of all columns of $Y_i^c$.

%After preprocessing (see section~\ref{sse:preprocessing}), the class covariance matrices are diagonal dominant. In addition, the background activity is modelled as a Gaussian white noise. Therefore, the background activity being dominant in relation to the signal of interest, the class-covariance matrices can be considered as diagonally dominant matrices. 

\subsubsection{The statistical model.}
\label{sss:model.details}

%We propose a model aiming at describing the within-class variability of the  %preprocessed signals. 
  
For a given participant, we consider $N^c$ trials per class, $c\in \{0,1\}$ being the class label. After preprocessing, we represent each trial $i$ as a vector $y_i^c \in \RR^{JK}$, where $J$ is the number of channels and  $K$  the number of retained wavelet and scaling coefficients.  In a given class $c$ and  according to the additivity assumption (see section~\ref{sss:model.definition}
above), $y_i^c$ ($i=1,\cdots, N^c$) is therefore written as 
\begin{equation}
\label{fo:MEM}
y_i^c = \mu^c + \Gamma^c b_i + \varepsilon_i\ ,
\end{equation}
where
\begin{itemize}
\item
$\mu^c\in\RR^{JK}$ is the mean vector of class $c$,
\item
$b_i\sim\cN(0,\tau^2)$ is a zero-mean Gaussian random variable with variance $\tau^2$,
  modelling the trial random effect,
\item
$\Gamma^c\in \RR^{JK}$ is a class-dependent coefficient vector which modulates
  the impact of the trial on each sensor and sample,
\item
$\varepsilon_i\sim\cN(0,\sigma^2 \mathbf{I}_{JK})$ is the residual part modelling the background activity where $\mathbf{I}_{JK}$ denotes the identity matrix of size 
$JK$. 
We further assume that the residual vector $\varepsilon_i$ and the
 trial random  effect $b_i$ are independent.
\end{itemize}

After specification of the coefficient vector $\Gamma^c$ (see~\sref{sss:design} below), the
model given in~\eref{fo:MEM} becomes a Linear Mixed Model (LMM), such that 
\begin{equation}
\label{loiY}
y_i^c |\Gamma^c \sim {\cal N}\left(\mu^c,V^c\right)\ ,
\qquad \mbox{where} \qquad V^c = \tau^2\Gamma^c \left(\Gamma^c\right)^\prime + \sigma^2 \mathbf{I}_{JK}\ ,
\end{equation}
(here $\left(\Gamma^c\right)^\prime$ denotes the matrix transpose of
$\Gamma^c$).

In summary,  each single-trial is decomposed into a fixed  part,
corresponding to the class-mean, a trial random effect $b_i$ modulated by the 
coefficient vector $\Gamma^c$, and a random term accounting
for the background activity.  
%Let us stress that the random effects are supposed class-dependent.

\smallskip
Let us note that the simplicity of the proposed model in wavelet domain  permits to obtain precise parameter estimates over a small number of trials. 
Indeed, after specifying the coefficient vectors $\Gamma^c$, we only have to estimate four parameters: the two mean vectors $\mu^c$, $c\in\{0,1\}$, and the variances $\tau^2$ and $\sigma^2$. 

%Some remarks are in order:

\begin{remark}\rm
\begin{enumerate}
%\item 
%modelling  directly  multisensor time series is hardly  for two main reasons. First, both sensors and time samples are
%strongly correlated. Therefore the background covariance matrix  is a full and large
%dimensional matrix, difficult to estimate and process numerically. Second, the
%high dimensionality of the data makes the problem computationally heavy.
%Preliminary transformations are hence necessary to obtain lower dimensional data,  that can be suitably modelled as in ~\ref{fo:MEM} using a diagonal matrix as  noise covariance matrix. 

\item
Since the class-mean is used as the fixed part,  the model described in~\eref{fo:MEM} 
reduces to a  mixed-effects ANOVA model~\cite{Searle1992}, where EEG-class is
the fixed effect factor and trial is the random effect factor. This allows
grounding the model on very solid foundations, and yields a more straightforward
interpretation of the results.

\item
In the case of the between-trial variability is negligible,  there is no random effect $\Gamma^c b_i$ in~\eref{fo:MEM} and the model corresponds to a simple linear one, as described in~\cite{blankertz:11}. 

\item
A related approach, that actually inspired the present work, was proposed by Huang~\etal in~\cite{Huang2008,Huang2009}. Our modelling, however, differs in three
major points. First, we consider a linear mixed model on wavelet domain more adapted to the assumption of  the noise decorrelation. Second, the fixed part is set to the class-mean and does not involve any projection. Third, the covariance matrix depends on channel, sample
and class while Huang~\etal propose dependence on trial, sample and class. We 
stress that our model is significantly simpler, avoids unnecessary approximations and requires less parameter estimations.

\item
Sticking to such a simple model is motivated by use cases where few trials are
available. This is the case under some experimental conditions where the number of
trials of interest is low, either because of the experimental design (\emph{e.g.}
probability bias), or because rare behaviour of the participants (\emph{e.g.}
errors in RT tasks). Another example is BCI type data, for which the training set has to be
limited to reasonable size, and to which high complexity models therefore can
hardly be fitted.
\end{enumerate}
\end{remark}

\subsubsection{Random part coefficient vector.}
\label{sss:design}
To completely specify the model in~\eref{fo:MEM}, the (class-dependent)
coefficient vectors $\Gamma^c\in \RR^{JK}$  have to be chosen. Let us note that
the components of these vectors modulate  the random effects $b_i$ in the same way for all trials in the same class.
$\Gamma^c$ is a design parameter to be fixed {\it a priori}. The model itself doesn't assume any specific form, the choice of $\Gamma^c$ is problem dependent, and usually relies on prior information. In the considered situation, such prior information is not available, the choice of $\Gamma^c$ was guided by preliminary  exploration of the data. We refer to~\sref{sse:gamma} for details.

%% Classification procedure 
\subsection{Classification procedure }

%The classification is based on posterior probabilities, i.e. each trial $i$ is
%classified according to the probability that it belongs to class $0$ or $1$
%(maximum a posteriori):  
%$$
%\max_{c \in \{0,1\}} \PP(\hbox{Class}=c\,|\,y_i)\,.
%$$  
%These  probabilities are  calculated by the Bayes formula.   
%Let us denote by $\pi_c$ the prior probability of the class $c$. 

Each single-trial $y_i\in\RR^{JK}$ to classify is assigned to a class $c$ according to the maximum \textit{a posteriori} calculated using the
Bayes formula:
\begin{equation}
\max_{c \in \{0,1\}} \PP(\hbox{Class}=c\,|\,y_i)\,.
\end{equation}

Based on the model proposed in~\eref{fo:MEM},  $y_i$ is
assumed to be distributed in each class $c$ according to the multivariate Gaussian distribution $\cN(\mu^c,V^c)$ where $\mu^c$ is the class-mean and $V^c$ is the class covariance matrix such that 
$V^c =\tau^2\Gamma^c \left({\Gamma^c}\right)' + \sigma^2 \mathbf{I}_{JK}$. 

 Under these
assumptions, the optimal classification is obtained with the following
quadratic discriminant function for each class $c$ :
\begin{equation}
\label{eq:discrimfunction}
g_c(y_i) = (y_i-\mu^c)^\prime ({V^c})^{-1} (y_i-\mu^c) -2\ln(p^c) + \ln\vert V^c \vert
\, ,
\end{equation}
which yields to the following binary classifier :
\begin{equation}\label{classif}
\delta(y_i) = g_0(y_i) - g_1(y_i)\ .
\end{equation}
%A simple decision rule is to assign the trial $y_i$ to class 0 if $\delta(y_i) < 0 $ and to class 1 otherwise. 
The trial $y_i$ is assigned to class 0 if $\delta(y_i) < 0 $, to class 1 otherwise. 

\begin{remark}\rm
Let us note that the resulting classifier is a particular case of the QDA classifier presented in~\sref{sse:LDAQDA}.
% In QDA, the covariance matrices are supposed class-dependent, while LDA
%assumes equal covariance matrices in each class. Indeed the classic QDA classifier is based on
%the following model :
%$$
%y_i= \mu^c + \eta_i^c\, ,
%$$
%where $\eta_i^c  \sim  \mathcal{N}(0,\Sigma^c)$.
%Usually, the semi-definite positive matrices $\Sigma^c$ are unknown and have to be
%estimated (i.e. $JK*(JK+1)/2$ coefficients must be estimated, which can be very difficult to achieve in practice). 
%
%If the trial random effect is removed in Model (\ref{MEM}), the covariance matrix $V^c$ is constant and the classifier $\delta$ is the same that the  LDA %classifier.  
In the situation considered here, the class-covariance matrix $\Sigma^c$ is
decomposed in the following simple form $ \tau^2\Gamma^c \left({\Gamma^c}\right)^\prime +
\sigma^2 \mathbf{I}_{JK} $. Therefore, after specifying the coefficient vector $\Gamma^c\in
\RR^{JK}$ for each class, only $\tau^2$ and $\sigma^2$ have to be estimated,
which makes QDA tractable in this situation.
\end{remark}
% However  the parameters $\mu^c$, $\tau^2$ and $\sigma^2$ are unknown. the parameter estimations are performed on a training step and the estimates are pluging in the expression (\ref{classif}).  

\subsection{A specific methodology for EEG single-trial classification}
\label{se:methodo}
We shortly summarize the main steps of the LMM-based classification
procedure (hereafter termed LMMC), which
consists in two stages: a training step to estimate the model parameters 
and the test step  to classify the EEG single-trials. For the reasons exposed
above, in many contexts as BCI, the training set shall be as small as possible to get
sufficiently precise estimates to correctly classify test set.
  
\subsubsection{Training Step.}
\begin{enumerate}
\item \textbf{Preprocessing.}\\ 
First whitening and time-domain reduction are performed through DWT. The $K$ selected multiscale coefficients are scaled to set the coefficient variability independent of the decomposition level. 
Then a spatial dimension reduction is performed. Spatial
  filters are estimated using a matrix-based Fisher's linear discriminant and 
only the $J$ most discriminant channels are selected.

\item \textbf{Wavelet-based modelling.}\\
Single-trials in the multiscale domain are  modelled as in~\eref{fo:MEM}, with the class-mean as a fixed component and the random component depending on class, channel and trial. 
In the latter, the class and channel dependence is given by the coefficient vectors $\Gamma^0$
and $\Gamma^1$ that must be specified a priori, according to~\eref{Gamma}
in this model.
   
\item \textbf{LMM parameter estimation.} \\
After fixing $\Gamma^c$ a priori, the complete parameter $\theta=(\mu^0,
\mu^1,\tau^2,\sigma^2 ) \in \RR^{2JK+2}$ of the LMM can be
estimated by likelihood-based methods using the assumption that $b$ and
$\varepsilon $ are independent and normally distributed. We use Restricted Maximum Likelihood (REML) method that provides unbiased variance
estimates (see \emph{e.g.}~\cite{Searle1992}). The estimates will be denoted by
$\hat\theta=(\hat\mu^0, \hat\mu^1,\hat\tau^2,\hat\sigma^2 )$.

\end{enumerate}

\subsubsection{Test Step.}
\label{sss:test}
\begin{enumerate}
\item \textbf{Preprocessing.} \\
The transformations are the same as in the training step. 

\item \textbf{Classification rule.} \\
The classifier $\hat\delta(y_i) $ is obtained from  the plug-in estimates of the  discriminant functions given in~\eref{eq:discrimfunction}. In the expressions~\eref{loiY} to~\eref{classif} the parameters $\mu^0$, $\mu^1$,$\tau^2$,$\sigma^2$ are replaced by their estimates computed during the training step. 

From the variance component estimates $\hat \tau^2$ and $\hat \sigma^2$,  we obtain the estimation of the class-covariance  matrix $V^c$ described in the model (\ref{loiY}) :
\begin{equation}
\label{Covariance.estimation}
\widehat V^c= \hat\tau^2 \Gamma^c(\Gamma^c)^\prime + \hat\sigma^2 \mathbf{I}_{JK}\ .
\end{equation}

\end{enumerate}

\subsection{Single-trial reconstruction }
\label{ss:estimatSI}
In addition, an estimate $\hat{b}_i$ can be obtained for each single-trial $i$ in class $c$ through the Henderson mixed model equations (see \emph{e.g.}~\cite{Searle1992}) :
\begin{equation}
\label{Henderson.equations1}
\hat b_i = \hat \tau^2 (\Gamma^c)'(\widehat V^c)^{-1}(y_i-\hat\mu^c)\ .
\end{equation}
This, in turns, yields an estimate for the signal of interest as follow :
\begin{equation}
\hat y_i^c = \hat\mu^c + \Gamma^c \hat b_i\,.
\label{eq:Yestimate}
\end{equation}

Notice that unlike Huang~\etal~\cite{Huang2008,Huang2009}, this estimation is a sub-product of the model allowing single-trial visualization and does not play any role in the classification procedure.

%\pagebreak
% Application

\section{Application to error potentials detection}
\label{se:application}

%***************************
% dataset - JNE

\subsection{ErrP Data and preprocessing}
\label{se:data}

We now  consider the  problem of modelling and detecting error potentials using the approach developed above on the dataset reported in~\cite{Roger2010}.
We start with a brief description of main aspects of the experiment that are relevant for the present study. Details on recordings and EEG data preprocessing are not exposed here since they had been largely detailed in~\cite{Roger2010} and are beyond the scope of the present paper.

\begin{figure}[h]
	\centering
		\includegraphics[width=0.40\textwidth]{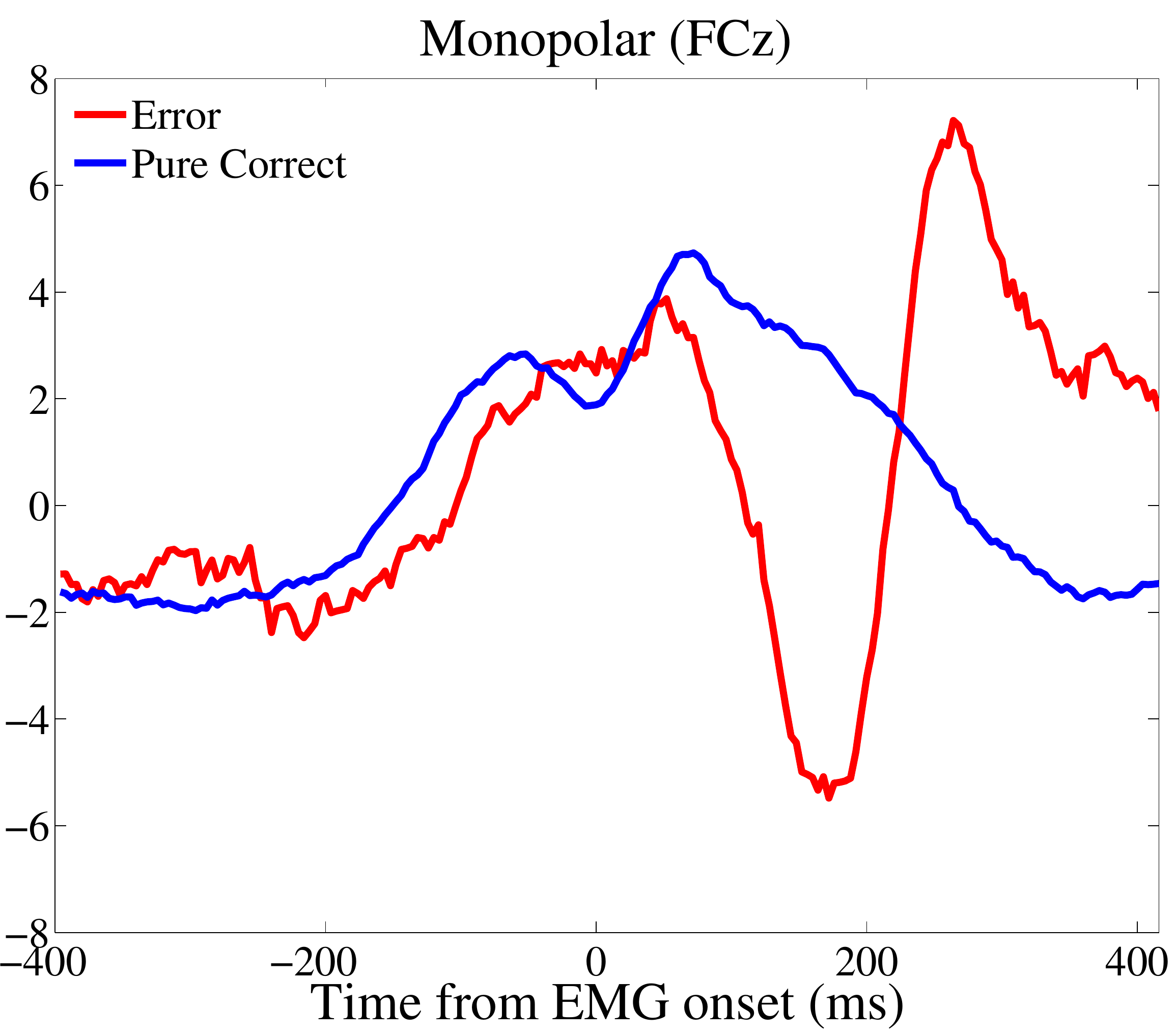}
	\caption{\label{FCz_Monop_Correct_Errors} \small Grand averages (FCz) of \textit{error} (red) and  \textit{correct} (blue) trials for monopolar data.}
\end{figure}

%***************************
\subsubsection{\it Experiment and signal acquisition.}
The dataset includes $10$ participants performing an Eriksen's flanker task \cite{Eriksen1974}.
Each trial consists in the identification of a central letter (target) embedded in a set of three letters. Participants were asked to respond to the target letter by pressing a button with either the left hand or right hand. The signal was recorded with $64$ scalp electrodes and after preprocessing (sampling to $1024$ Hz and artifacts removal), data were downsampled to $256$ Hz.

The selected trials were segmented into epochs of $800$ ms from $-400$ ms to $+400$ ms, were zero corresponds to the electromyogram (EMG) onset that triggered the response. Each dataset in the training step correspond to a \textit{three-dimensional array} as we consider $N^c$ trials in both classes $c=0$ (correct trials) and $c=1$ (errors) over $64$ electrodes and $204$ time-points.

%***************************
\subsubsection{\it Error/Correct trials classification: an unbalanced dataset.}
For each participant, we consider two categories of trials: \textit{error} and \textit{correct}. In this current application we do not take into acount the third class discussed in~ \cite{Roger2010} which corresponds to partial errors. In~\fref{FCz_Monop_Correct_Errors}, differences on monopolar data between \textit{error} and \textit{correct} are observable on grand averages (over trials and subjects).  For errors, a clear negativity is observable around $150$ ms after the EMG onset (zero of time) whereas on correct trials a large positive wave occurs just after EMG onset, but no negativity is visible~\footnote{On correct trials, Current Source Density (CSD) analysis has revealed that a negative activity similar to the one observed on errors exists, but with a much smaller amplitude \cite{Roger2010}. In the present context, since discrimination was the main goal, we did not resort to CSD analysis for maximizing the difference between correct and errors. This is why such a small negative wave is not observable on correct trials in this case}.

The number of trials per class differs greatly for each participant. In table~\ref{tab:NbTrials}, the total number of trials for error and correct responses is given. The percentage of error trials approximately ranges from $2\%$ to $12\%$, the dataset is therefore highly unbalanced, errors  forming the minority class.
% As a prior information based on all participants, we may assume that the errors represent approximately  $10\%$ of the total number of trials.
  The main goal here is to achieve {\it single-trial classification} on this unbalanced dataset.

For an illustrative purpose throughout the section, results are presented for participant $A$. Results concerning the other participants  are given in Supplementary Data.

%***************************
%\subsubsection{\it Data dimension  }
%The selected trials were segmented into epochs of $500$ ms from $-100$ ms to $+400$ ms, were zero corresponds to the EMG onset that triggered the response. Each dataset in the training step correspond to a three-dimensions matrix. Thus, we consider $N^c$ trials in both classes $c=0$ (correct trials) and $c=1$ (errors) over $64$ electrodes and $128$ time-points. . As a prior information based on all the participants, we may assume that the erros represent $10\%$ of the total number of trials in each dataset.

\begin{table}[h]
\caption{\label{tab:NbTrials}Total number of trials per participant for error and correct responses.}
\begin{indented}
\item[]
\begin{tabular}{@{} l c c c c c c c c c c}
\br
& \multicolumn{10}{c}{\textbf{Participant}} \\ \ms
\cline{2-11}
\ms
 & A & B & C & D & E & F & G & H & I & J\\
\mr
Error & 130 & 105 & 43 & 39 & 18 & 28 & 94 & 145 & 100 & 63 \\
Correct & 1376 & 1575 & 1467 & 1735 & 690 & 759 & 1844 & 1238 & 1167 & 907 \\
\br
\end{tabular}
\end{indented}
\end{table}

\subsubsection{Preprocessing.}

The following two-step preprocessing for spatial and temporal decorrelation and dimension reduction is applied to the ErrP data.

\paragraph{Decorrelation and time-domain reduction.}
DWT was performed using a Daubechies filter D6 (see \cite{Daubechies92ten,Mallat08wavelet,Vetterli95wavelets} for details) with $5$ decomposition levels. Given the sampling frequency, those 5 levels correspond to the following frequency bands: $64-128 Hz$, $32-64Hz$, $16-32Hz$, $8-16Hz$ and $4-8Hz$ respectively, corresponding to wavelet coefficients, the lowest frequency band $(0-4 Hz$) being encoded in scaling coefficients (see section~\ref{ssse:wavelets}).

Prior to DWT, signals were first extended using zero-padding.
Doing so, we obtain a set of signals with appropriate length (for the DWT
implementation which we use~\cite{BuckheitWavelab}, it is convenient that the
signal length be a power of two). Zero-padding is also an extension method which
permits to avoid misinterpretation about the behaviour of signals beyond
boundaries.
\begin{remark}\rm
The choice of D6 is the result of a tradeoff between localization and smoothness for the resulting wavelet. Wavelets with larger support generate more important boundary effects (we recall that in the dataset under consideration, signals are epoched, so that boundary effects have to be accounted for). Shorter filters yield wavelets with lower smoothness, so that the adjusted signals can also lack smoothness. For example, adjusted signals obtained using the Haar wavelet are discontinuous, which we didn't find satisfactory, while adjusted signals obtained with D4 are non differentiable. Further tests were made using Daubechies filters with various lengths. Corresponding results (which can be found in Supplementary data) are fully consistent with the results presented here.
\end{remark}

\begin{figure}[h]
	\centering
		\includegraphics[width=1\textwidth]{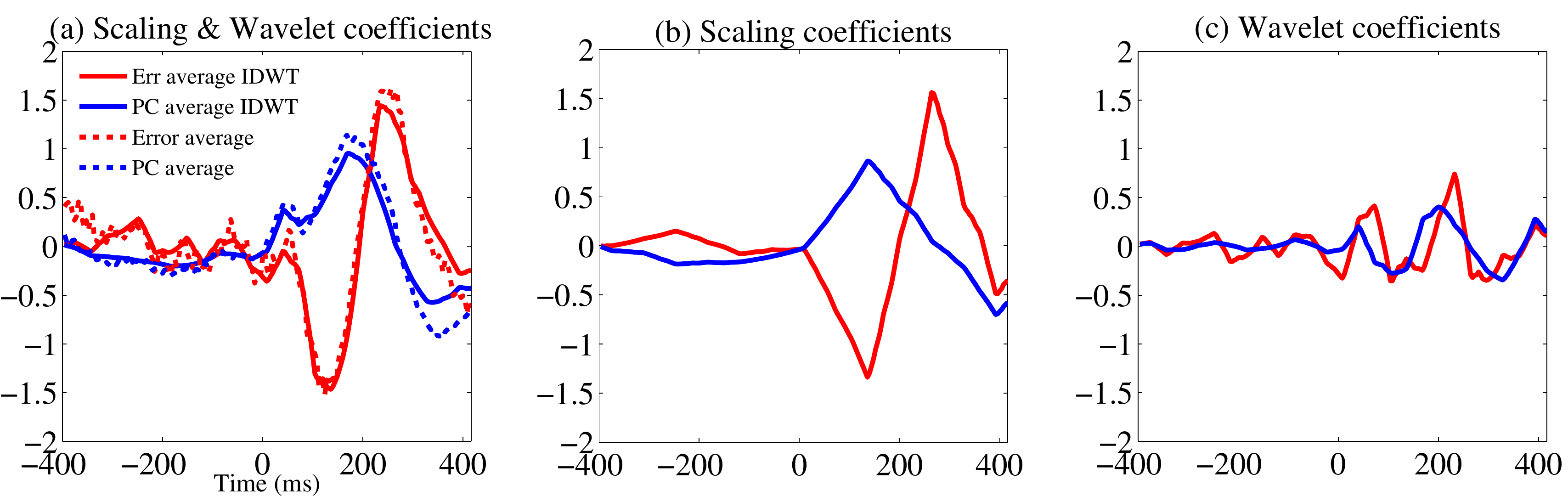}
	\caption{\label{fig:illustDWT} Illustration of the relevance of DWT and multiscale coefficients selection.
	 Figures (a), (b) and (c) represent error and correct responses averages over FCz. (a) Error (red) and correct responses (blue) averages
	 obtained from the $24$ selected coefficients projected back to time domain and compared with time-courses averages (dashed lines).
	 Partial reconstructions based on  scaling (b) and wavelet coefficients (c) are also represented separately.
	 %as they play different role in the DWT: scaling coefficients correspond to the waveform while wavelet coefficients correspond to details. }
	 }
\end{figure}

Second, dimension reduction was achieved by removing the first three decomposition levels,
that contain low amplitude high frequency phenomena in the signals. The truncation of the EEG
wavelet coefficients sequence to $2$ decomposition levels amounts to
a projection onto an appropriate subspace \cite{Abramovich2000} that keeps the
relevant information in the remaining wavelet and scaling coefficients.

Finally, keeping in mind that the signals were zero-padded, boundary wavelet coefficients that originate from the zero-padding are excluded from the statistical analysis, as they turn out to exhibit a behaviour different from the relevant ones. More precisely, coefficients likely to be sensitive to boundary effects were not taken into account, namely the 2 boundary scaling coefficients, the 2 boundary wavelet coefficients at coarsest scale, and the 4 boundary wavelet coefficients at finer scale. After this step, $K=24$ coefficients are selected: $12+6$ wavelet coefficients for decomposition levels $4$ and $5$ respectively and $6$ scaling coefficients.

\Fref{fig:illustDWT} illustrates the relevance of multiscale coefficients selection. The $24$ coefficients remaining
after DWT were projected back to the temporal space using an inverse DWT (IDWT). As seen in~\fref{fig:illustDWT}(a), errors and
correct responses averages obtained from the selected coefficients are very similar to the $204$ time-points averages.
In addition, one can see that scaling coefficients (b) capture the general waveform while wavelet coefficients concentrate
the details (c).

In the following, the selected multiscale coefficients are referred to as \textit{multiscale features}.

\begin{figure}[h]
	\centering
		\includegraphics[width=1\textwidth]{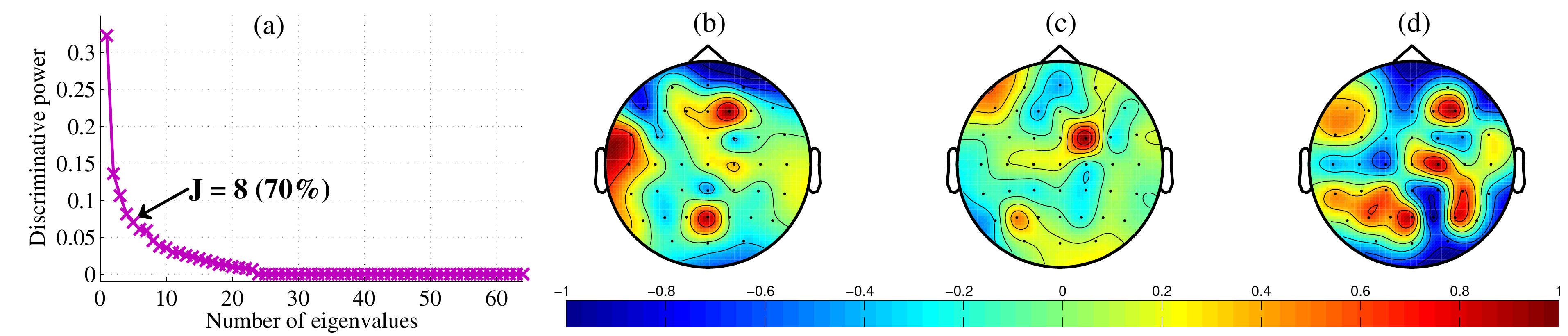}
	\caption{\label{fig:illusSpatialFilt1} Spatial filtering results for participant A. (a) Using a cut-off of $70\%$ of
	discriminative power, $J=8$ filters are selected based on the screeplot of eigenvalues. (b), (c), (d) display the three first selected filters, ordered by decreasing order of discriminative power. }
\end{figure}

%\begin{figure}[h]
%	\centering
%		\includegraphics[width=1\textwidth]{Illustration_SpatialReduc_1.pdf}
%	\caption{\label{fig:illusSpatialFilt1} Spatial filtering results for participant A. (a) Using a cut-off of $70\%$ of
%	discriminative power, $J=8$ filters are selected based on the screeplot of eigenvalues. (b), (c), (d) display the three first selected filters, ordered by decreasing order of discriminative power. }
%\end{figure}

\paragraph{Spatial filtering.}
Using the matrix-based Fisher's linear discriminant proposed in section~\ref{sss:SpatialFilt}, spatial filtering consists in identifying the most discriminant filters to separate errors and correct responses.
The number $J$ of selected filters depends on the cumulative percentage of discriminative power, fixed to $70\%$ in the present study for all subjects.
This choice is somewhat arbitrary. We have chosen to stick to a very simple criterion, but other criteria could have been used as well. For example, Kaiser's rule has been tested on our dataset, and results were not satisfactory in the sense that the number of selected channels was larger (and the computational load was increased), without improvement of the classification rates.

\begin{figure}[h]
	\centering
		\includegraphics[width=1\textwidth]{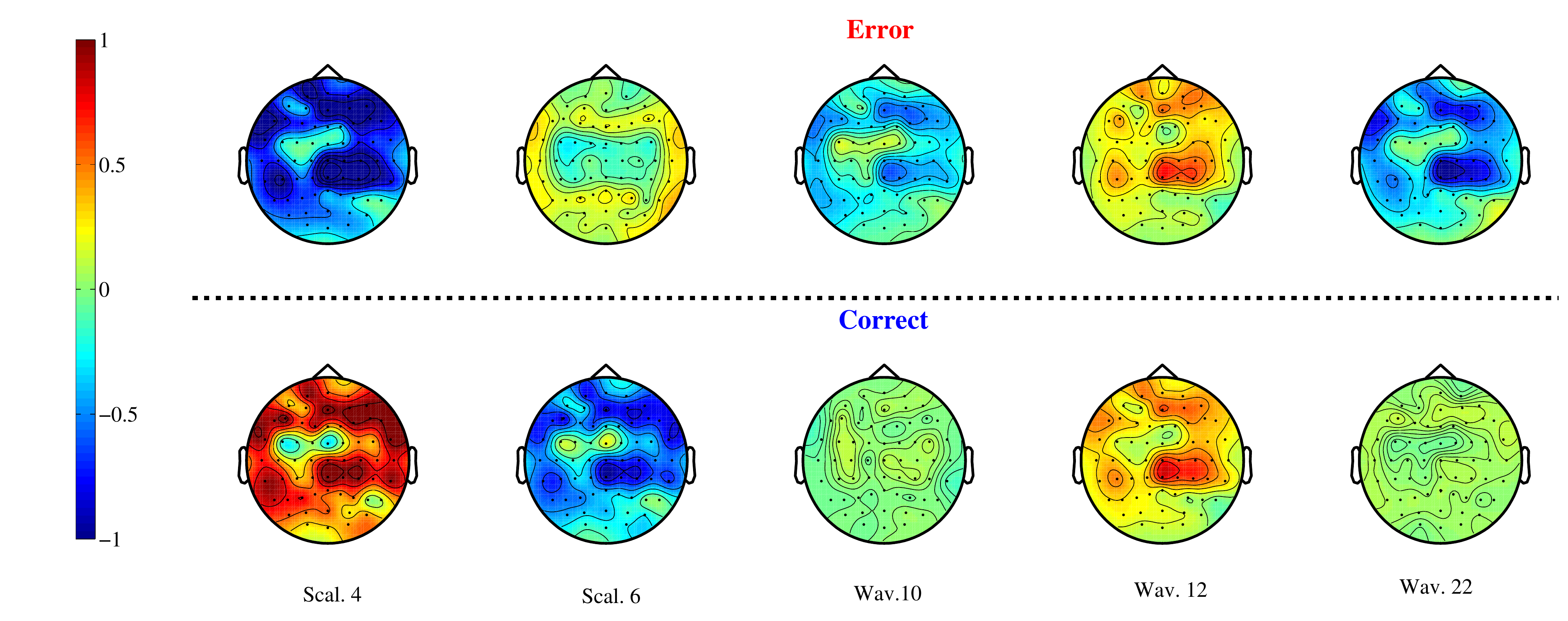}
	\caption{\label{fig:SpatialSignatures_SA} Topographical maps of spatial signatures of five multiscale features averaged across trials. }
\end{figure}

\Fref{fig:illusSpatialFilt1}(a) represents the screeplot, which plots the ordered eigenvalues, associated with the $64$ electrodes and
out of which only $8$ filters are needed to concentrate $70\%$ of the discriminative power.
\Fref{fig:illusSpatialFilt1} (b), (c), (d) correspond to topographical maps of the first three spatial filters i.e. the weights of the electrodes in the corresponding projections.
%As their interpretation is not straightforward, spatial features are also presented hereafter (see~\fref{fig:illusSpatialFilt2}) in order to
%complete the spatial analysis.
The contributions of multiscale features in each class  can be visualized using their spatial signature defined as their   backprojection onto the sensor space: for each multiscale index, the pseudo-inverse of the spatial filter matrix is applied to the  vector of the corresponding coefficient values at all channels. For the sake of illustration, we display in \fref{fig:SpatialSignatures_SA} the spatial signatures of some multiscale features averaged across trials.

Here we have chosen to display five features corresponding to scaling coefficients $4$ and $6$ and to wavelet coefficient $10$, $12$ and $22$. These specific multiscale coefficients were chosen for their relevance for \textit{error}, \textit{correct} or both classes: on average they appear as emergent, large magnitude coefficients (see \fref{fig:Hyp_SA} below). For example, scaling coefficient $4$ has the largest average magnitude in both classes and was found to be highly discriminant. This  behaviour is reflected by the corresponding spatial signatures (defined above)  which are strongly negatively correlated.
Scaling coefficient $6$ is relevant  only for the correct class and so is its spatial signature. Wavelet coefficients $10$ and $22$ are only relevant for error whereas  the coefficient $12$ is not discriminant. In agreement with a large literature on ErrP, the spatial signatures for the relevant multiscale features mainly loads on fronto-central sensors.

%\begin{figure}[h]
%	\centering
%		\includegraphics[width=1\textwidth]{Illustration_SpatialReduc_2.pdf}
%	\caption{\label{fig:illusSpatialFilt2} Topographical maps of mean spatial features.
%	For both classes, wavelet and scaling coefficients are dissociated. The maps of spatial features were obtained
%	by first projecting multiscale coefficients onto the space spanned by the $8$ spatial filters, then by recovering filtered signals in the sensor space using the pseudo-inverse of $\left[ u_1,\ u_2,\ \ldots,\ u_8\right]$ and finally averaging across scaling and wavelet coefficients.}
%\end{figure}

%\begin{figure}[h]
%	\centering
%		\includegraphics[width=1\textwidth]{SpatialSignatures_SA.pdf}
%	\caption{\label{fig:SpatialSignatures_SA} Topographical maps of mean spatial signatures. For both classes, wavelet and scaling coefficients are dissociated. For error we select the $4th$ scaling coefficient (a) and the $22th$ wavelet coefficient (b).
%	For correct we select the $4th$ scaling coefficient (c) and the $12th$ wavelet coefficient (d).
%	 The maps of mean spatial signatures were obtained by first projecting the selected multiscale coefficients onto the space spanned by the $8$ spatial filters, then by recovering filtered signals in the sensor space using the pseudo-inverse of $\left[ u_1,\ u_2,\ \ldots,\ u_8\right]$ and finally averaging across trials. }
%\end{figure}&

\subsection{Modelling ErrP single-trials}

After applying spatial and temporal filters, the extracted features from ErrP data are
decorrelated and dimensions are significantly reduced.
We now consider multichannel trials in the wavelet domain, such that $Y_i^c \in \RR^{K\times J}$,
with $K = 24$ and $J \ll M$ (for the training set example $J=8$, see~\fref{fig:illusSpatialFilt1}). Modelling ErrP single-trials is based on the columnwise vectorized trial $y_i^c$.

\subsubsection{Testing class-covariance matrices inequality.}
\label{sse:Boxtest}
\begin{figure}[h]
	\centering
		\includegraphics[width=0.8\textwidth]{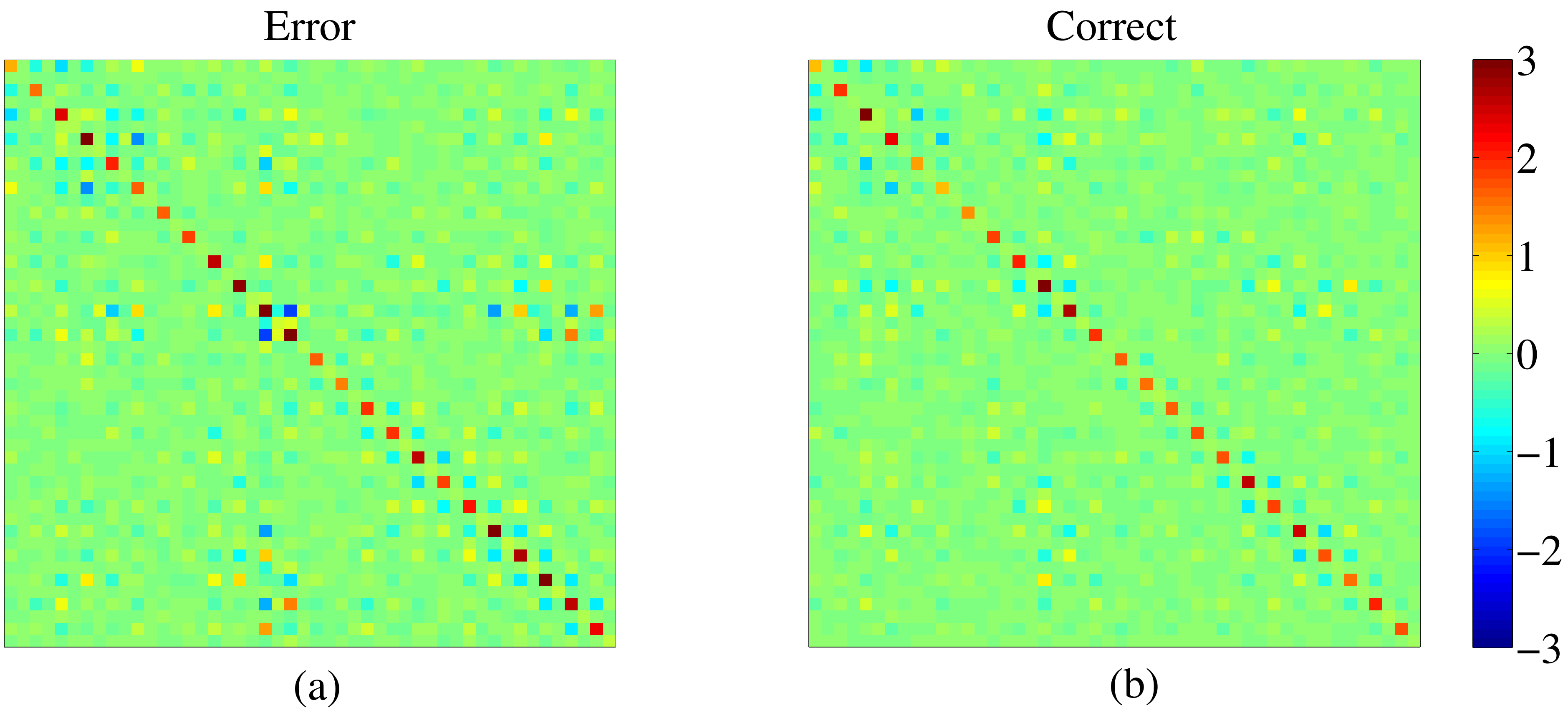}
	\caption{\label{fig:Cov_KJ} Representation of $48 \times 48$ sample class-covariance matrices for 24 multiscale coefficients and 2 channels : {\it error} class   (a) and {\it correct} class (b).  }
\end{figure}

%Class-covariance inequality is tested using a statistical test~\cite{Anderson2003}.
 We display in~\fref{fig:Cov_KJ} the  sample covariance matrix for each class (\textit{error}  and \textit{correct}), calculated after preprocessing.  For simplicity the figure is limited to the first two channels. In addition for the relevance of the visual comparison, the two  matrices have been computed on datasets of equal size (actually maximal possible size, to ensure maximum precision).  Visual inspection reveals different diagonals and noticeable non-diagonal terms in the sample covariance matrix of the  \textit{error}  class. The significance of the difference between these two matrices  is  evaluated quantitatively using Box's M test~\cite{Anderson2003}. The latter  clearly rejects the  class-covariance matrices equality null hypothesis (p-value $\ll 10^{-6}$).

Importantly, this conclusion differs from the claim of~\cite{blankertz:11}, who did not see any noticeable  difference between sample class-covariance matrices (on a different EEG dataset). Although such a finding may be dataset dependent, we stress that for the dataset considered here, the significant difference results from time and space dimension reductions, which yield  an  important  denoising.
In the absence of  dimension reductions (for example raw data), the sample covariance matrices are generally singular and the Box's M test cannot be performed.

%\begin{figure}[h]
%	\centering
%		\includegraphics[width=0.8\textwidth]{CovMatrix_KJ.pdf}
%	\caption{\label{fig:Cov_KJ} Representation of $48 \times 48$ class scatter matrices for 24 multiscale coefficients and 2 channels : error class   (a) and correct class (b).  }
%\end{figure}

%\begin{table}[h]
%\caption{\label{tab:MBoxTest} Box's M test for homogeneity of the two covariance matrices $\Sigma^{Err}$ and $\Sigma^{PC}$.
%Box's M tests the null hypothesis that the observed two scatter matrices are equal. The Box's M test statistic (3273.05) is significant with 48 degrees of freedom (df) (p-value $<< 10^{-6}$). $\Sigma^{Err}$ and $\Sigma^{PC}$ are significantly different. }
%\begin{indented}
%\item[]
%\begin{tabular}{@{} l l l l l l}
%\br
%MBox & Chi-sqr & g & p & df & p-value \\
%\mr
%3273.05 & 2859.74 & 2 & 48 & 1176 & $<< 10^{-6}$ \\
%\br
%\end{tabular}
%\end{indented}
%\end{table}

\subsubsection{Choice of the random part coefficient vector $\Gamma^c$.}
\label{sse:gamma}
\begin{figure}
	\centering
		\includegraphics[width=0.9\textwidth]{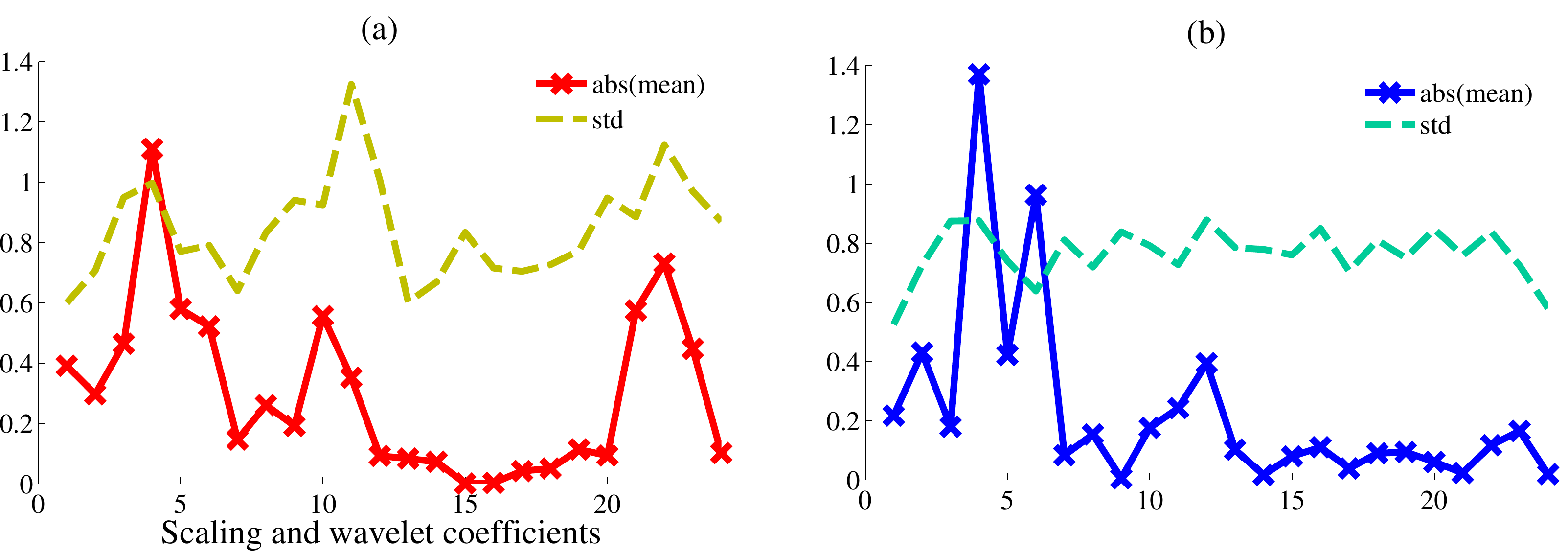}
	\caption{	\label{fig:Hyp_SA} Illustration of the monotonic relationship between  amplitude and  variability in ErrP-related
	EEG signals: mean
in absolute value ({\it abs(mean)}) and to the standard devation ({\it std}) both calculated across all trials of the first channel.  (a): \textit{error} ; (b): \textit{correct}.}
\end{figure}

As mentioned in~\sref{sss:design}, the choice of $\Gamma^c$ is problem dependent. In our application no prior information was available that could help specifying the form of $\Gamma^c$, we thus relied on preliminary data exploration.
We plot in~\fref{fig:Hyp_SA} the mean in absolute value and the standard deviation computed over all trials for participant A on the first channel in the two classes. This reveals a monotonic relationship between the amplitude and variability of the signal of interest, with an additional offset effect. The latter can be interpreted as originating from random noise $\varepsilon$ in~\eref{fo:MEM}.
Similar results are obtained for the other participants and channels.

In the present work, we take the simplest relationship, namely a linear one, by setting the vector $\Gamma^c$ to the average over trials in class $c$\,:
\begin{equation}
\label{Gamma}
\Gamma^c=\frac{1}{N^c}\sum_{i=1}^{N^c} y_i^c\ .
\end{equation}

\subsubsection{Variance components estimation.}

We now explore one of the main assumptions of our study, namely the single-trial signal of interest can be written as the sum of
two components: a class-dependent fixed term (the class-mean) plus a trial-dependent random term (the so-called
random effect).
The hypothesis to be tested is  the significance of  the random effect variance  $\tau^2$. A negative answer would indicate
that a random effect is not needed. In that case, a classical linear model as proposed in~\cite{blankertz:11}
would be more appropriate.

\begin{figure}
	\centering
		\includegraphics[width=1\textwidth]{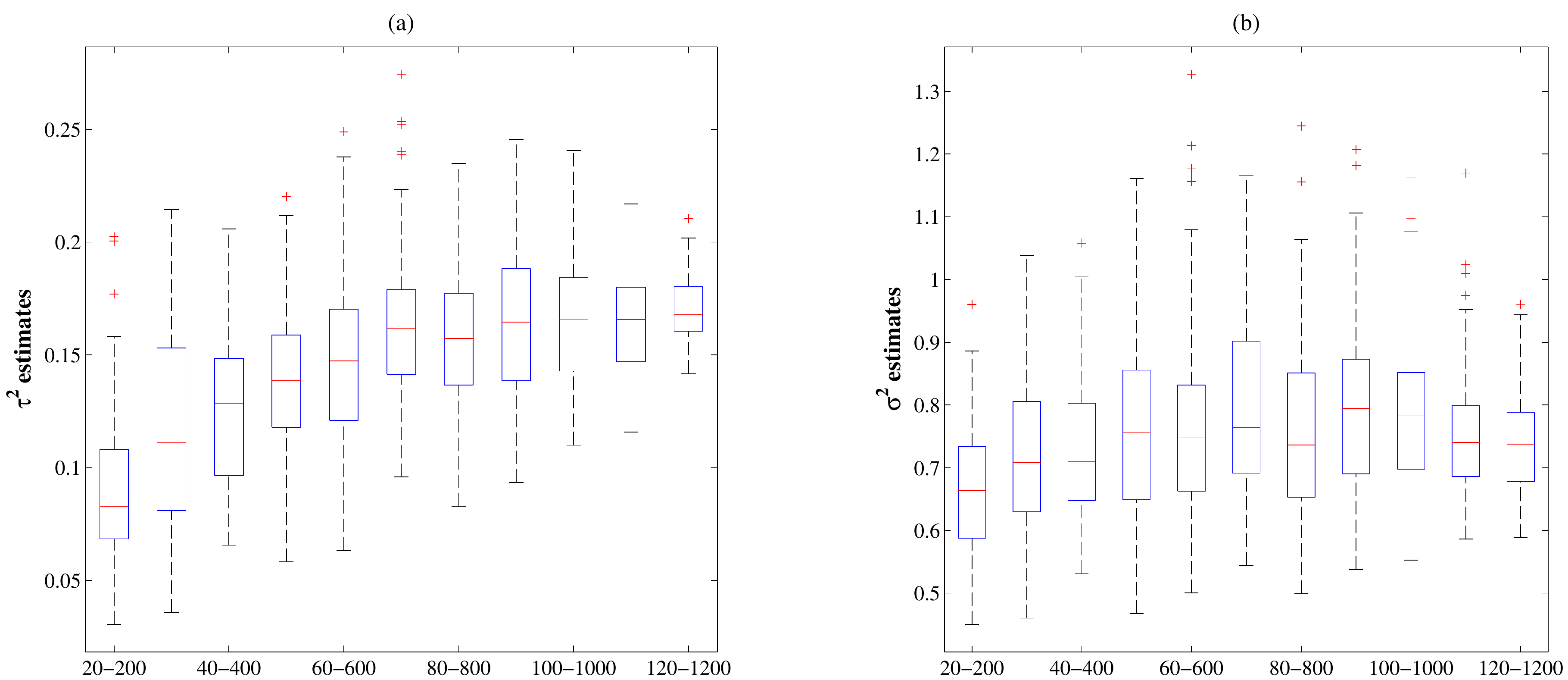}
	\caption{	\label{fig:Tau2_Sigma2} Covariance components estimation. (a) Random effect variance $\tau^2$ estimates.
	(b) Residual variance $\sigma^2$ estimates. Each boxplot displays the series   of 100 parameter estimates for  different training sample sizes ($10\%$ error trials, $90\%$ correct trials).
	 \textit{The central mark of  boxplot is the median (second quartile), the bottom and top of the box are  the first and third quartiles, the whiskers extend to the most extreme values not considered outliers, and outliers are plotted individually}.  }
\end{figure}

The implementation of the mixed model has been done using the MATLAB function {\it mixed.m} written by Witkovsk\'y~\cite{Witk2002} (more details  can be found in Supplementary Data).
REML estimates for variance components $\tau^2$ and $\sigma^2$ are displayed in~\fref{fig:Tau2_Sigma2}. Each boxplot corresponds to a different sample size ($10\%$ error trials, $90\%$ correct trials) and summarizes the distribution of estimates performed on  $100$ different training sets. We note that the magnitude of the residual variance estimates $\hat{\sigma}^2$ is larger than that of $\hat{\tau}^2$. Globally the dispersion of the estimates decreases as a function of the sample size, nevertheless the figure shows that good estimates can be obtained even with small sample sizes.
In \fref{fig:Tau2_Sigma2} (a) the quartiles of the $\hat\tau^2$ distribution tend to increase slightly with the sample size and  minimum estimates rise above zero. Using a z-test we test that the random effect variance is strictly positive in average for each training sample size (p-value $< 10^{-3}$).

% Non-null $\hat{\tau}^2$ (Wald test) indicates that there is a significant single-trial variability in both classes.
% (see~\fref{fig:Tau2_Sigma2} (a)), justifying the proposed mixed model~\eref{fo:MEM}.
%Let us note that for smaller sample size ($N^1 = 10$ and $N^0 = 100$), the number of parameters to estimate is larger than
%the number of observation ($2 \times KJ +2 > N$), leading to poorly estimated covariance parameters.

%Using the asymptotic Wald test~\cite{Wald1943} for testing the significance of $\tau^2$,
%we can conclude with a critical level of $1\%$ that the random effect variance is, in average and for each training sample size, significantly different from zero.

\subsection{Classification results}
\label{se:classif.res}
\subsubsection{Performance evaluation.}

In the case of unbalanced classes, the performance evaluation methodology has to be chosen carefully as the majority class may shadow the behaviour of the minority class. This problem has been discussed in several  works (see \emph{e.g.} ~\cite{Weiss2001,Weiss2004,Menardi2010} and references therein).
In this paper, we use various performance   measures  for comparing unbalanced EEG dataset classifiers.
In particular, we focus on  class-dependent evaluations.

\begin{itemize}
\item[-] \textit{Confusion matrix}. In a binary classification, given a classifier and a test sample, four outcomes are possible:
true positive, false positive, true negative  and false negative.
For instance, if the test sample is positive and it is classified as negative, it is counted as a false negative.
%; if it is classified as negative, . If the instance is negative and it is classified as negative, it is counted as a
%true negative (TN); if it is classified as positive, it is counted as a false positive (FP).
For a test dataset, the number of occurences of the four outcomes are  respectively denoted by (TP), (FP), (TN) and (FN), and form the $2\times 2$ confusion matrix given in table \ref{tab:confMatrix}.

\begin{table}[h]
\caption{\label{tab:confMatrix}Confusion matrix for the ErrP  classification problem.}
\begin{indented}
\item[]\begin{tabular}{@{}l l l l}
 \br
 & & \multicolumn{2}{c}{predicted} \\
 & &  $corret$ & $error$ \\
 actual & $correct$ & TN & FP \\
 & $error$ & FN & TP \\
 \br
\end{tabular}
\end{indented}
\end{table}

%From the confusion matrix, two main indicators can be defined:
%\begin{itemize}
%\item The True positive rate (TPR), or sensitivity:
%\begin{equation}
%TPR = \frac{TP}{(FP+TP)}
%\end{equation}
%\item The True negative rate (TNR) or specificity:
%\begin{equation}
%FNR = \frac{TN}{FN+TN}
%\end{equation}
%\end{itemize}
%
%The Receiver operating characteristic (ROC) curve is constructed by plotting the False Positive Rate (FPR = 1- TNR = 1 - Specificity) and the TPR for
%different thresholds. Roc curve is one of the most frequently used tools for evaluating the accuracy of a classifier in the
%presence of unbalanced classes as a trade-off between minimizing FPR and maximizing TPR can be done varying the classification threshold.
%
%The classifier performs as better the larger the area underlying the curve (AUC) is.

\item[-] \textit{Good classification rate.} In the particular case of unbalanced data, the good classification rate must be calculated
for each class independently. Indeed, a global good classification rate may lead to misleading results. For example, in our problem, where the errors only represent $10\%$ of the data, the naive classification strategy allocating all testing data to the majority class would automatically achieve a good classification rate of $90\%$. However the classifier is obviously irrelevant for detecting $error$ trials.

\item[-] \textit{Pierce's Skill Score.} Given the
confusion matrix, the good prevision rate $H = TP/(FP+TP)$ and the false alarm rate $F = FP/(TN+FP)$ for the $error$ class lead to the so-called Pierce score
\begin{equation}
\label{eq:PSS}
PSS = H - F \,.
\end{equation}
By construction, $PSS\in \left[ -1;+1 \right]$.  In our context the closer $PSS$ is to $+1$, the better the classifier is for the detection of  \textit{error} trials.
\end{itemize}

%***************************
\subsubsection{ErrP detection.}
\label{sse:ClassifResult}

For the considered dataset the number of \textit{error} trials ($N^1$) is much smaller than the number of \textit{correct}  trials ($N^0$). For performance evaluation,
unbalanced training sets of increasing size are generated, all based on the \textit{a priori} $10\%$ of \textit{errors}.
For each participant and each randomly
 drawn unbalanced training set (namely, $\{N^1=20, \ N^0=200\}\,, \{N^1=30, \ N^0=300\}\,, \ldots$), the test set is composed of all remaining data, therefore its size differs. For each participant this splitting is performed   $100$ times and classifiers performance are recorded.

\begin{figure}[h]
\includegraphics[scale=0.45]{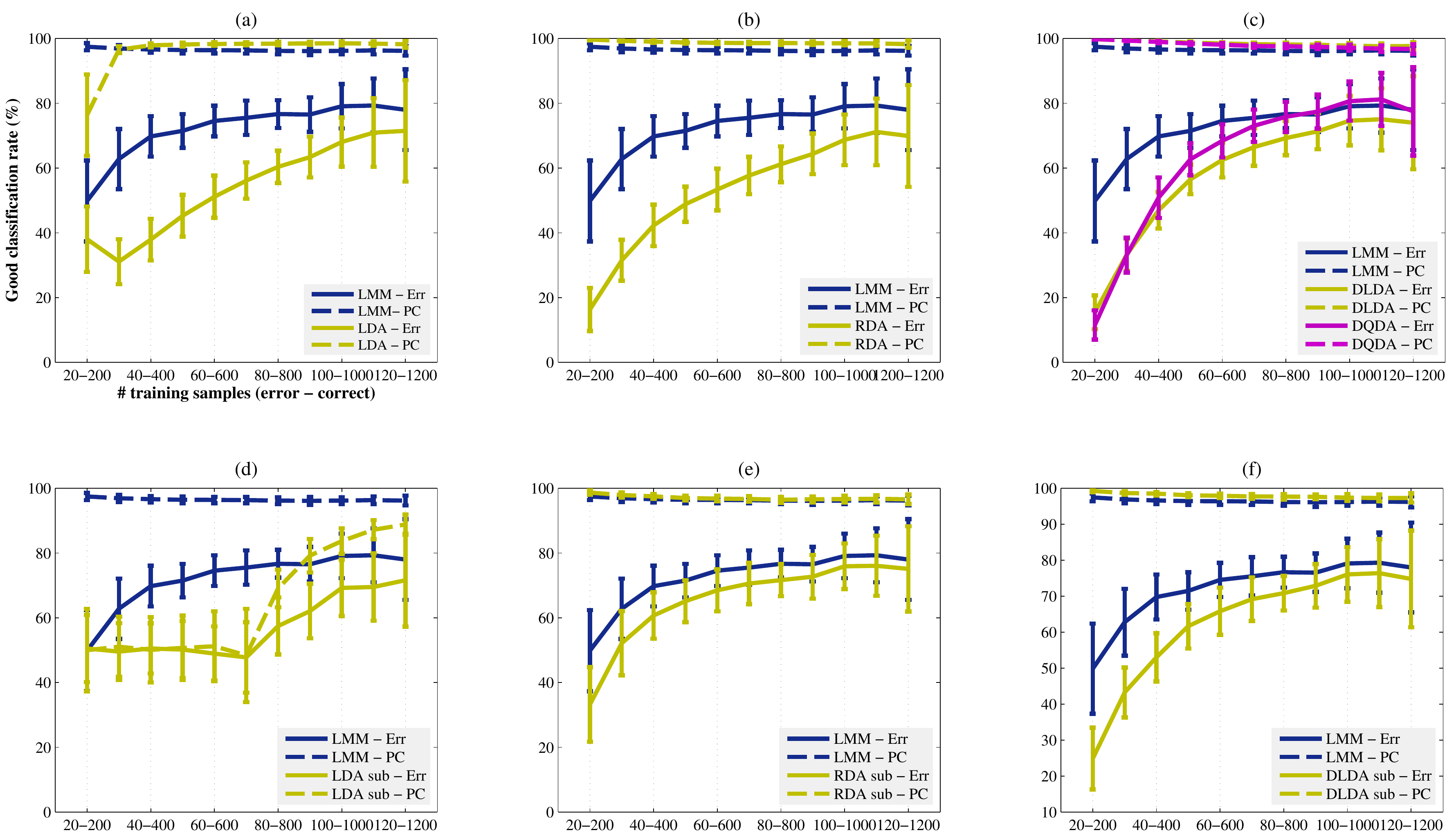}
\caption{\label{fig:Classif_SA}  Good classification rates for different training sample sizes ($error-correct$). Results correspond to good classification rates averaged over $100$ iterations for each training sample size. Vertical bars represent   standard deviations (std). Results are given for \textit{error}  (plain lines) and \textit{correct} (dotted lines) test trials.  LMM classifier is compared with three different classifiers on the unbalanced and re-balanced datasets. (a), (b), (c) display results in the unbalanced case ($10\%$ \textit{error} trials) for classic LDA (LDA), Regularized LDA (RDA) and Diagonal LDA and QDA (DLDA $\&$ DQDA) respectively.
(d), (e), (f) display results of the LDA, RLDA and DLDA classifiers after re-balancing datasets by subsampling the {\it correct} class. }
\end{figure}

The proposed method (named LMM hereafter) is compared with several classifiers:
classical linear discriminant analysis (LDA), regularized method (regularized LDA - RDA - as proposed in~\cite{blankertz:11}) and diagonal discriminant analysis (diagonal LDA (DLDA) and diagonal QDA (DQDA)). In addition, LMM is also compared to LDA, RDA and DLDA classifiers after  sub-sampling the  majority class to balance training sets. The classic QDA classifier cannot be used, because the sample covariance matrix in  {\it error} class is very often singular in the considered range of sample sizes.      Corresponding  results are displayed in \fref{fig:Classif_SA}. Let us stress that the error bars represent the standard deviation of the good classification rates distribution. Since the training and test sets are complementary, when the size of training set increases, the number of single-trial signals used to evaluate the good classification rates decreases, and consequently the error bar increases.  These error bars only measure the dispersion of the estimates and cannot be used for quantitative pairwise comparison  (see below).

Clearly LMM greatly outperforms each method in terms of \textit{error} detection in all situations.
This result is particularly spectacular for small training set which we interpret as follows.
The simplicity of the underlying model (which involves very few parameters) and its relevance for the considered dataset yield accurate and robust parameter estimates even for small sample size.
In addition, our method performs equivalently to other methods to detect \textit{correct}  trials which correspond to the majority class.
Even with small sample size, classifiers detect \textit{correct} trials with a good classification rate close to $100\%$.

We provide in~\tref{tab:PSS_SA_sign} quantitative classifiers comparison based on Pierce's Skill Score (PSS) defined in~\eref{eq:PSS}. For each training set, PSS is computed for each classifier. LMM is compared with each other classifier (OC) using Wilcoxon signed rank test (a paired non-parametric statistical test, see~\cite{Gibbons2010,Hollander2014}) based on the average difference $PSS_{LMM}-PSS_{OC}$.

For participant A, LMM outperforms all the other considered classifiers very significantly except DQDA from the sample size ($80-800$). Indeed the DQDA  performances are equivalent to or significantly better  than the LMM ones  for the largest  sample sizes.
Quantitative results for all participants   can be found in Supplementary Data. The conclusions drawn   for participant $A$ are confirmed in most configurations but should be modulated as follows.
In most situations LMM performs consistently better than the other classifiers. This in particular true for  small training sets namely when the number of errors is in the range (20, 30, 40, 50), where the superiority is supported by statistical tests (with only two exceptions where results are equivalent).
 These results can be explained by the preprocessing which enforces the data decorrelation and are coherent with the results of Box's M test (see section~\ref{sse:Boxtest}).

The proposed method is therefore a valuable alternative to linear discriminant techniques which involves quadratic decision rule avoiding the difficulty of usual shortcomings of classical QDA.

\begin{table}[h]
\caption{\label{tab:PSS_SA_sign} \emph{Participant A}: Classifiers comparison based on PSS. LMM performances are compared with other classifiers using the Wilcoxon signed rank test. For each pair of classifiers LMM-other classifier,  the difference between the two mean PSS is given and p-value order of magnitude is provided between parentheses. }
\begin{indented}
%\noindent
\lineup
\item[]
\begin{tabular}{@{}lllll }
\br
& \multicolumn{4}{c}{\textbf{Classifiers comparison}} \\ \ms
\cline{2-5}
\ms
$N^1$-$N^0$ & LMM-LDA & LMM-RDA  & LMM-DLDA & LMM-DQDA \\
\mr
20-200 & $0.14 \ (10^{-12})$ & $0.36 \ (10^{-18})$ & $0.37 \ (10^{-18})$ & $0.41 \ (10^{-18}) $ \\
30-300 & $0.34\ (10^{-18})$ & $0.34 \ (10^{-18})$ & $0.32 \ (10^{-18})$ & $0.32 \ (10^{-18})$ \\
40-400 & $0.35 \ (10^{-18})$ & $0.30 \ (10^{-18})$ & $0.25 \ (10^{-18})$ & $0.21 \ (10^{-18})$  \\
50-500 & $0.28 \ (10^{-18})$ & $0.25 \ (10^{-18})$ & $0.16 \ (10^{-18})$ & $0.09 \ (10^{-16})$  \\
60-600 & $0.25 \ (10^{-18})$ & $0.23 \ (10^{-18})$ & $0.13 \ (10^{-18})$ & $0.07 \ (10^{-13})$  \\
70-700 & $0.21 \ (10^{-18})$ & $0.19 \ (10^{-18})$ & $0.10 \ (10^{-18})$ & $0.03 \ (10^{-4}) $ \\
80-800 & $0.18 \ (10^{-18})$ & $0.17\ (10^{-18})$ & $0.08 \ (10^{-17})$ & $0.009 \ (0.43)$  \\
90-900 & $0.14 \ (10^{-18})$ & $0.13 \ (10^{-18})$ & $0.06 \ (10^{-13})$ & $-0.01 \ (10^{-3})$  \\
100-1000 & $0.12 \ (10^{-17})$ & $0.11 \ (10^{-17})$ & $0.05 \ (10^{-10})$  & $-0.02 \ (10^{-4})$ \\
110-1100 & $0.09\ (10^{-12})$ & $0.09 \ (10^{-13})$ & $0.04 \ (10^{-7})$  & $-0.02 \ (10^{-4})$ \\
120-1200 & $0.07\ (10^{-4})$ & $0.08 \ (10^{-6})$ & $0.04 \ (10^{-2})$  & $0.005 \ (0.23)$ \\
\br
\end{tabular}
\end{indented}
\end{table}

\begin{remark}\rm
As mentioned above, the correct classification rate depends heavily on the quality of the estimation of the covariance matrices. In our case, the latter are strongly simplified, and the issue is the quality of the estimation of the mean vectors $\mu^c$ and the variances $\tau^2$ and $\sigma^2$. Thanks to pre-processing and dimension reduction, variances are estimated from $KJ(N^0+N^1)$ samples, while mean vectors $\mu^c$ are estimated from $N^c$ samples. Hence the quality of the classification is mainly driven by the quality of the estimation of the mean vectors.
Since we consider unbalanced situations, the difficult class is the minority class (in our case the {\it error} class) for which approximately 30 samples or more are needed to get more than 60\% correct classifications. The other class being ten times bigger, the corresponding classification rate is close to 100\%, and the global classification rate is satisfactory.
\end{remark}

Similar results using different Daubechies filters are presented in Supplementary Data. Whatever the filter, the proposed method consistently outperforms concurrent approaches in terms of classification, best results being obtained when the Haar filter is used.

\subsection{Back to time courses: adjusted single-trials}

\begin{figure}[h]
\includegraphics[scale=0.45]{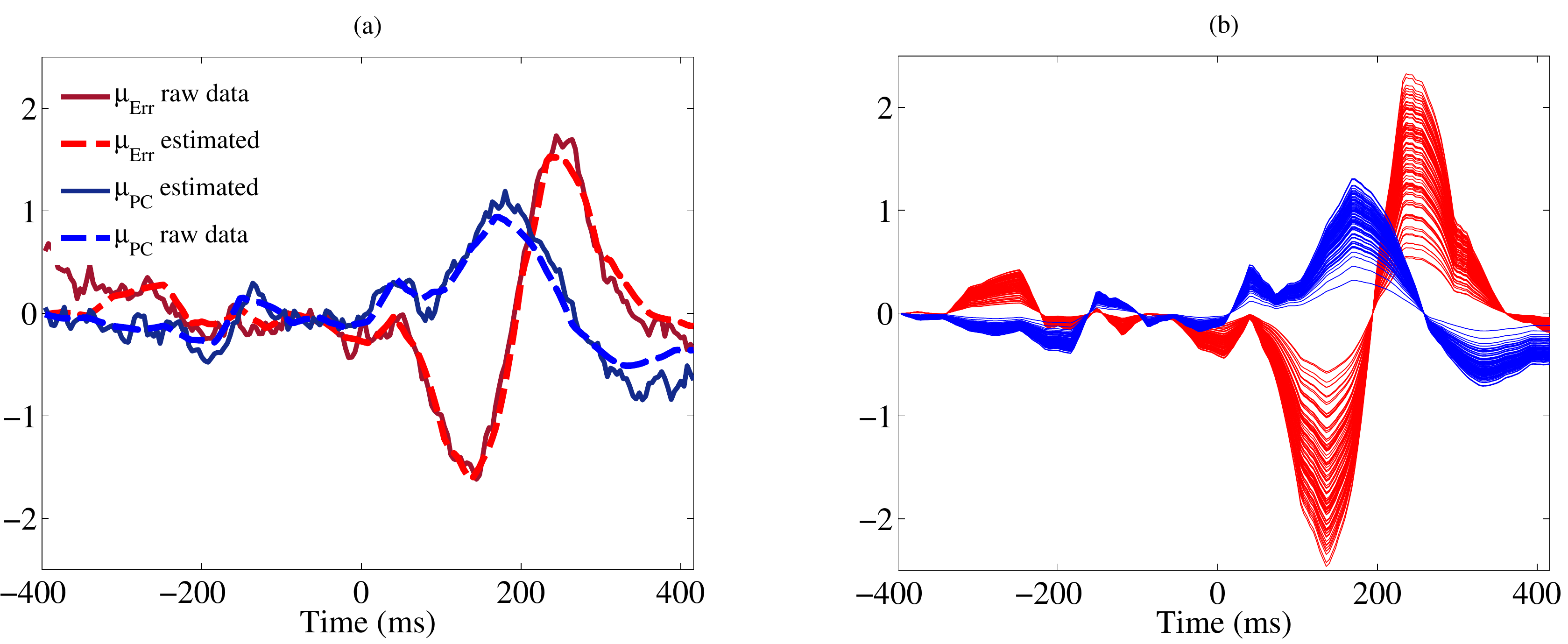}
\caption{\label{fig:adjustedtrials} Illustration of single-trial reconstruction. (a) \textit{Error} and \textit{correct} trial averages on FCz for participant A. Dashed curves:   averages  across raw trials in both classes;  full curves:  averages across adjusted trials. (b) Adjusted single-trials for \textit{Err} (red) and  \textit{PC}(blue) classes on FCz. }
\end{figure}

As described in subsection \ref{ss:estimatSI}, for each single-trial an estimate of the signal of interest can be obtained according to~(\ref{eq:Yestimate}). After back projection onto sensor space (using a  Moore-Penrose pseudo-inverse) and inverse DWT we obtain time courses called  \textit{adjusted single-trials}. Examples of adjusted single-trials (electrode FCz) are displayed in~\fref{fig:adjustedtrials}. These plots are obtained using the D6 Daubechies filter. Adjusted single-trials obtained using other Daubechies filters are displayed in Supplementary Data.

The left hand plot (a) represents the raw  signals averaged over trials (within each class) together with the averages of adjusted single-trials. As can be seen for each class the two averages are very close to each other for participant A.  For other participants, this similarity is less striking but still present. This shows that neither dimension reductions nor modelling have strongly affected the average information.  Although not surprising and even expected, the recovery of the global shape of the average is a first necessary step before looking at single-trial reconstruction. It appears that the modelling indeed captures the signal of interest and filters out the background activity.

 As largely reported in the literature (see \cite{falkenstein:00} for an overview), in the \textit{error} class, the estimated  averages as well as the adjusted single-trials show a clear negative wave around
$150$ ms after EMG onset followed by a positive one. Such a negative wave is not observable in the \textit{correct} class, as commonly reported with monopolar recordings.
%The estimated \textit{ErrP} averages nicely recover the shape of the raw averages, with smoother time course.
%More importantly, as the proposed method specifically models single-trials variability, the estimation of the variance components allows a \textit{reconstruction}
%of the signal of interest for each single-trial through the LMM estimates.
%\Fref{fig:adjustedtrials}(b) illustrates single-trial estimation using~(\ref{eq:Yestimate}).

The added value of the model is the explicit description and the numerical quantification of  the inter-trial variability. The latter appears clearly in the right hand plot of \fref{fig:adjustedtrials}(b) where adjusted single-trials are displayed.
The proposed method therefore allows one to reveal the amplitude variation  of the ErrP: while some errors induce a very large activity, some others  present virtually no response. The same figure is displayed for each participant in Supplementary Data.

Although speculative at this point, this could reveal differences in sensitivity to the errors across trials, an hypothesis that now becomes testable thanks to the proposed method.

%\pagebreak

% Discussion - JNE

\section{Discussion}
\label{se:discussion}

%In this section we highlight the main contributions of the present work and
%%we propose several applications which may be derived from the current one.

We first highlight the  key points and main contributions of our method and then discuss other possible  applications of this work. 
%***************************

\subsection{The impact of time-domain decorrelation}
Within the proposed procedure, we would like to emphasize that the introduction of DWT constitutes a key step for
the analysis and the classification of single-trials.
%Not only it allows time-domain decorrelation and dimension reduction, but it also
%has important consequences for the modelling and testing phases.
Considering multiscale coefficients instead of time-point values first gives 
a simplified representation of data, but it also has important consequences for both modelling and classification.  
Indeed, decorrelation and dimension reduction lead to small size, diagonal dominant covariance matrices, easier to estimate. Consequently, in the wavelet domain  single-trials can be modelled by simple linear mixed model  with a small number of unknown parameters, allowing robust estimates from small unbalanced training set and good classification results.

%Finally, the LMM build in the wavelet domain characterized, with a good precision, single-trials
%with a small number of parameters to estimate and provide good classification results
%in the unbalanced case. 

\subsection{Fine structure of the model}
Mixed model provides a general and flexible approach to analyze single-trial EEG signals. It allows one to include both fixed and random effects within the same analysis. In the case of EEG datasets it appears natural to consider trial effects as random effects since they can be seen as originating from a random selection from a much larger set.  Furthermore, it is of direct relevance to separate inter and intra-trial variability.  However for each problem and dataset of interest, the main difficulty lies  in the choice of the dimension of the random effect vector, its class-dependence and the covariance structure.   

%Several related models have to be considered and some form of model selection must be used %\textcolor{red}{Je ne suis pas sur de comprendre ce point\dots}. 
% <<<<<<< .mine
For ErrP detection problem,  the training sample size being small, we choose the simplest possible linear mixed model: class as fixed effect (class-average), one dimensional random effect vector (class-independent) modulated by a class-dependent coefficient vector $\Gamma^c$ plus a white noise.  The choice of $\Gamma^c$ depends on the dataset and can be done using prior information or exploratory data  analysis. In our application the observed relationship between amplitude and variability led us to set $\Gamma^c$ equal to the average over trials in each class. 

% In the time domain linear mixed model proposed in \cite{Huang2008} the random part coefficient vector depends  of trials,and it was found by using some LDA.

\subsection{Relevant classification results}
%In this work we are interested to compare  methods  based on  linear models with the same fixed part (the class-average) with different variability modellings in the  unbalanced case.
% <<<<<<< .mine
In section~\ref{se:classif.res}, we provided  a systematic comparison of related classification techniques, all based upon linear models, involving the same fixed part (class-average) and various approaches to model variability in  the unbalanced situation.
All methods were applied after the same preprocessing step, that enforces diagonal dominance for covariance matrices. Consequently, methods that can exploit such diagonal dominance properties are particularly relevant. As expected,  results show that they significantly outperform classic LDA or QDA in terms of classification.
Moreover our classification results highlight two important points. The first one is that our procedure is particularly efficient when the size of the training sample is small, emphazising the relevance of the proposed variability modelling through the linear mixed model. 
Secondly, beyond a certain sample size, namely when the number of observations permits to estimate precisely the diagonal of the covariance matrix in the minority class, diagonal QDA outperforms all considered methods. That point is coherent with the results of the Box's M test of equality of class-covariance matrices. For all participants, when the number of trials is sufficient to calculate the test statistic, Box's M test rejects the hypothesis of equality (p-value $\leq 10^{-3}$). Unlike the balanced case where LDA generally outperforms QDA even when class-covariance matrices are different, diagonal QDA gives better results than diagonal LDA. In the unbalanced case, we stress that it is preferable to take into account the difference in variability of both classes. However it is well known  that the classifier performance depends on the  accuracy of the estimate of parameters involved in the procedure, and to obtain accurate parameter estimates the number of samples must be  much  larger than the number of parameters. Therefore when the sample size is small, modelling the intra-class variability is well suitable. In our model, in addition to the class average estimates, only two parameters are needed to estimate the two class-covariance matrices.

\subsection{Possible applications}
Being able to exploit the single-trial %information 
M/EEG component would open a new window on the dynamics of brain processes, and
the range of applications is wide. Although some might be obvious, let us focus on some areas where the proposed approach might be especially useful. 

\subsubsection{BCI.} In the present application, we considered a discriminant framework. In this context, BCI naturally becomes a straightforward extension for the proposed LMMC procedure. Indeed, classification results, presented in~\sref{sse:ClassifResult}, illustrate the performance of single-trial variability modelling on classifying $error$ vs $correct$ trials. 
Importantly, the total variance explicit modelling, combined with spatial and temporal dimension reductions, allows robust parameters estimates, even when based on a limited training dataset. Sticking to a small training dataset is essential to keep the BCI calibration session as short as possible, a condition that can be reached  thanks to the simplicity of the proposed model.
The procedure, as described above, is not restricted to error potential estimate, however, and can easily be extended to other BCI protocols. 
Indeed, a strength of LMMC method is the modularity and  interchangeability of the preprocessing steps.
Concerning the spatial dimension reduction, well established spatial filtering methods have been developed for specific BCI tasks: for example Common Spatial Pattern (CSP) are especially suited spatial filters for motor imagery~\cite{Sanelli2010} whereas xDAWN has been developed for P300 data~\cite{Rivet2009}. Those methods can easily be implemented in the LMMC procedure while leaving the modelling step almost unchanged.
The same holds for temporal dimension reduction. In the current application, to
get rid of temporal correlation and for temporal dimension reduction, we used
wavelet transform which is well suited to extract transient phenomena such as
ERP's (ErrP, P300, etc.). However, the choice of the basis can depend on the
data to analyze, and for more oscillatory signals such as the ones recorded in
motor imagery (like $\mu$ and $\beta$ rhythms, see~\cite{McFarland2000}),
other transforms such as time-frequency transforms (MDCT, STFT,
see~\cite{Mahanta2012}) or adaptive transforms~\cite{Vautrin2009,Barbieri13optimal} might be more suitable. 
Again, changing the transform leaves the variance modelling steps almost unchanged, as long as the transform is invertible.

\subsubsection{Single-trial extraction in cognitive neurosciences.}
In many experimental situations, being able to extract the single-trial amplitude of brain activity would open a new window on the dynamics of brain processes. One example is the temporal evolution of brain activity during learning: across repetitions of the learning situation, some brain regions might become more active, while some other might disengage from the task. Relying on averaging, at best, dramatically reduces the analysable temporal dynamic of the learning process. For example, while no clear learning effect could be evidenced on the raw Auditory Evoked Potentials (AEP) in rats, extracting the single-trial component (through denoisig in this case) allowed to regress learning curves on the trial--by--trial responses, 
and show that the amplitude of some components of the AEP reduces  with learning while other remain unchanged \cite{quiroga:02}. 
% Extracting the single-trial amplitude allows studying, at a very fine grain, the temporal evolution of brain response (see \emph{e.g.}~\cite{quiroga:02} for an example in rats). 
Averaging also prevents analyzing correlations between brain activities and behaviour across trials. For example, RT or psychophysical judgements are known to be highly variable even for the very same stimulation. The variability in RT have long be known to provide essential information to put into test different model of information processing \cite{luce:86}. Unfortunately, the equivalent variability is normally not accessible for EEG/MEG brain responses, preventing the use of similar constraints on models based on brain activity (see \emph{e.g.} \cite{Gerson2005}, for trial-to-trial variability to constraint RT models). The LMM approach  furnishes a way to recover this variability (at least in amplitude) and may help to correlate brain activities to behaviour. Further applications will probe what type of constraint such variability can bring for model testing.

Another key interest of the proposed approach is the ability to better analyze unbalanced data, which is a quite common situation. Indeed, from a statistical point of view, LMM allows increasing the robustness of the estimates from small datasets and hence reducing the risk of type II error (false negative), which is a common problem in small and noisy dataset. It may also help to decrease the number of necessary trials to reach a given statistical power, which can be of tremendous interest on some populations (children, patients, etc.).
%\emph{[est-ce que c'est correct de dire celà???}]

One may wonder, however, how general this methodology can be and whether it would apply with similar success on other brain activities which might not be as stable as the ErrP. Although we do not have any empirical response, formal analysis of the model allows speculating a bit. In the current modelling, the class average plays a central role. One can therefore anticipate that the quality of the  modelling will largely be driven by the quality of the average. As a consequence, as long as the average emerges from the background noise, that is, as long as there is an ERP, it should be possible to apply the model. For example, it should be possible to apply the method on robust components, such as the standard visual evoked potentials (N1, P2, N2 etc.). This will be investigated on future research.

% CONCLUSIONS - JNE

\section{Conclusions and future works}
\label{se:conclusion}
In the present work we propose a single-trial classification procedure (LMMC)   based on an explicit model of the signal variability in the  wavelet domain.   Each  single-trial signal is assumed to be the sum of two components:
a background activity and a signal of interest. The latter can again be decomposed into the sum of a fixed part (the class-mean) and a random part  
that models the deviation of each single-trial from the class-mean.  This hypothesis is formalized mathematically as a Gaussian linear mixed model (LMM).   This model generalizes the Gaussian linear model given in \cite{blankertz:11} which was adapted to situations where between-trial variability is  negligible. An important step is the discrete wavelet transform in the preprocessing.  On the one hand, this transformation allows one to reduce the time dimension and on the other hand, to consider a very simple model where class-covariance matrices are characterized by only few parameters (to be estimated). The advantage of such a model is to yield 
robust estimates of parameters from a small dataset.

This model is applied to a classification problem, namely separating correct from erroneous responses in a RT task. As expected, the proposed classification procedure  is particularly efficient to detect EEG Error Potentials   in the unbalanced case when the sample size of the minority class is small. Moreover these results  validate the chosen modelling. 
  
%In this present paper, we propose an explicit modeling of the signal variability to allow single-trial analysis and classification. Each raw signal is assumed to be the sum of two components:
%a background activity and a signal of interest. The latter can again be decomposed into the sum of a fixed part (the class-mean) and a random part  
%that models the deviation of each single-trial to the class-mean.  This hypothesis is formalised mathematically as a Gaussian LMM. This model generalises the Gaussian linear model given in \cite{blankertz:11} which corresponds to the situation when the between-trial variability is  negligible.  
%The proposed model is applied to a classification problem, namely separating correct from erroneous responses in a Reaction Time (RT) task
%and provide significant results in the case of small and unbalanced dataset. 

In addition, the proposed procedure allows the extraction of relevant and discriminant information from EEG signals. 
As a matter of fact model parameters estimates provide quantitative measures of the between-trial variability in each class in addition to the average behaviour. 
In the specific context of error potential detection, the proposed model
allows one to recover  single-trial signal of interest  and more interestingly reveals the trial-by-trial amplitude variation of ErrP's. 

%{\it Compared to more standard techniques (see e.g. \cite{quiroga:00,wang:07}) that would first perform some denoising and then estimate the variability of denoised trials, LMM features a global approach to estimate the different sources of variability and to extract the relevant signal} \textcolor{blue}{je propose d'enlever  cela ou de le reformuler car nous aussi   on enleve  du bruit avant la modelisation en supprimant les hautes frequences apres la DWT}.

Several directions might be interesting to investigate further and we list a few of these below. 

\noindent First, the choice that was made for random part coefficient vector
$\Gamma^c$ explicitly assumes a linear relationship between the signal of
interest's amplitude and variability. However, although the existence of a monotonic dependence between mean and variance seems a reasonable assumption, the relationship needs not be linear and is likely to be more complex (for example, class dependent).
%it sounds reasonable to assume that the variance increases monotonically with the mean, this increase needs not be linear.
Further investigating and characterizing the link between mean and variance
should improve the choice of the random component and hence the
modelling performance. 

\noindent Another improvement would be to better take into account the cubic structure of M/EEG data ({\textit i.e} space $\times$ time $\times$ trials). Decoupling spatial and temporal features extraction~\cite{Spinnato2014} could lead to a more complex form for the random effects design matrix and hence to a spatio-temporal modelling of between-trial variability.
%The introduction of the spatio-temporal structure in the random part would allow observing different sources of variability for each single-trial. 

Comparing the EEG error potentials across participants, not only on this dataset but in the literature, see \emph{e.g.} \cite[figure 3]{Roger2010}, suggests large similarities between participants, but also some differences, both in terms of topography and time courses.
%\noindent \textcolor{red}{[\emph{je propose:}] Comparing the EEG error potentials across participants, not only on this dataset but in the litterature, see \emph{e.g.} \cite[figure 3]{Roger2010}, suggests similarities large between participants, but also some differences, both in terms of topography and time courses.}\textcolor{blue}{B: je pense que c'est bien comme \c ca}.
%
%\noindent  [ Finally, comparing \textcolor{blue}{the results obtained for the} different participants (see figure \ref{FCz_Monop_Correct_Errors} \textcolor{blue}{ce n'est pas tres clair en quoi   la figure 1 apporte des indications sur un comportement commun des sujets, il faudrait un peu plus expliciter les choses à mon avis et indiquer dans nos resultats ce qui laisse a penser cela. D'autre part je n'ai pas la figure 3...} {, and \cite[figure 3]{Roger2010}) suggests a large commonality between participants, both in term of time course and of topography. ]
%%Following~\cite{Fazli2011}, 
One might therefore consider incorporating also the between participant variability within a more  sophisticated mixed model, which would then model the deviation of the participant (in time and in topography) to the mean behaviour. The model would include common settings and parameters specific to each subject. From a statistical point of view, the introduction of such common settings would presumably increase the robustness of the estimates from small datasets.
%\textcolor{red}{and hence to reduce the risk of type x error, which is a common problem in small, and hence noisy,
%
%\textit{From a statistical point of view the expected gain is to increase the robustness of the estimates from small datasets \textcolor{red}{and hence to reduce the risk of type x error, which is a common problem in small, and hence noisy, }}
%\textcolor{blue}{Boris penses-tu qu'il faut   ajouter quelque chose au point de vue neuro? } \textcolor{red}{je propose de remonter cet argument, car il est déjà valide sans modélisation inter-sujet (fin de discussion). Je propose ici un autre argument, plus sur l'inter sujets.}
This would in particular allow a more robust estimate of the single participant topography and time course. The question as to whether such differences in topography is related to fine anatomical differences \cite{Procyk2014} would become easier to test, especially when few trials are available.

%One might consider that the between participants variability could also be modeled as a fixed term, which would be the grand average, plus a random effects, specific to each participant, which would model the deviation of the participant (in time and in topography) to the mean (ref for across participants mixed-effects). The two types of mixed model (random effects on participants and on trials) might even be combined, so that each individual trial would be decomposed into a random (trial specific) and a fixed effect (across trials), which would in turn be decomposed into a random (participant specific) and a fixed effect (across participant)

%Finally, comparing the different participants
%(see Figs.~\ref{fig:GrandAverage_ErrP_AllSubjects}
%and~\ref{fig:GrandAverage_PC_AllSubjects}, as well as
%Fig.~\ref{FCz_Monop_Correct_Errors}, and \cite[Fig. 3]{Roger2010})
%suggests a significant commonality between participants.
%% both in terms of time course and topography. 
%Following~\cite{Fazli2011}, one might therefore consider accounting for the
%between participant variability through a more sophisticated mixed model, which
%would then model the deviation of the participant (in time and in topography) to
%the mean behaviour.

%***************************

% Acknowledgements
\ack
The authors are grateful to the anonymous reviewers, whose valuable questions and comments have contributed to significantly improve the quality of the manuscript.\\
This work was supported by the ANR project CO-ADAPT (ANR-09-EMER-002-05). J. Spinnato's work is funded by a PhD grant from R\'egion Provence-Alpes-C\^ote d'Azur, France.
B. Burle is supported by a European Research Council under the European Community's Seventh Framework Program (FP7/2007-2013 Grant Agreement no. 241077).
This work has been carried out in the framework of the Labex Archim\`ede (ANR-11-LABX-0033) and of the A*MIDEX project (ANR-11-IDEX-0001-02), funded by the ``Investissements d'Avenir" French Government programme managed by the French National Research Agency (ANR).
%*******************************
% Biblio
%*******************************
\bibliographystyle{abbrv}
\bibliography{Biblio_JNE}
\end{document}